\documentclass[12pt]{article}
\usepackage{amsmath,amssymb,latexsym,graphicx}
\usepackage[usenames]{color}

\newcommand{\emp}{{\scriptstyle\rm empty}}

\begin{document}
\begin{flushright}\small
 \begin{minipage}[b]{3cm}
  {YITP-06-11\\KUNS-2019\\OIQP-06-04\\hep-th/0603242}
 \end{minipage}
\end{flushright}

\begin{center}
 {\Large\bf Dirac Sea and Hole Theory for Bosons I}\\[1em]
 {\large\it --- A new formulation of quantum field theories ---}\\[2em]
{%
Yoshinobu \textsc{Habara}$^{1}$, %
Yukinori \textsc{Nagatani}$^{2}$,\\%
Holger B. \textsc{Nielsen}$^{3}$ %
and Masao \textsc{Ninomiya}$^{4}$\footnotemark}\\[1em]
{\it%
$^1$Department of Physics, Graduate School of Science,\\%
Kyoto University, Kyoto 606-8502, Japan\\[0.5em]%
$^2$Okayama Institute for Quantum Physics,\\%
Kyoyama-cho 1-9, Okayama-city 700-0015, Japan\\[0.5em]%
$^3$Niels Bohr Institute, University of Copenhagen,\\%
17 Blegdamsvej Copenhagen \o, Denmark\\[0.5em]%
$^4$Yukawa Institute for Theoretical Physics,\\%
Kyoto University, Kyoto 606-8502, Japan%
}
\end{center}

\begin{abstract}
{%
Bosonic formulation of the negative energy sea,
so called Dirac sea, is proposed
by constructing a hole theory for bosons
as a new formulation of the second quantization of bosonic fields.
The original idea of Dirac sea for fermions,
where the vacuum state is considered as a state completely filled
by fermions of negative energy
and holes in the sea are identified as anti-particles,
is extended to boson case in a consistent manner.
The bosonic vacuum consists of a sea
filled by negative energy bosonic states,
while physical probabilities become always positive definite.
We introduce a method of the double harmonic oscillator
to formulate the hole theory of bosons.
Our formulation is also applicable to supersymmetric field theory.
The sea for supersymmetric theories has an explicit supersymmetry.
We suggest applications of our formulations
to the anomaly theories and the string theories.
}
\end{abstract}

\footnotetext{Also working at Okayama Institute for Quantum Physics, Kyoyama-cho 1-9, Okayama-city 700-0015, Japan}

\section{Introduction}

In any relativistic quantum field theory
there inevitably appear positive energy solutions
as well as negative energy ones~\cite{Weinberg:1995mt}.
These two kinds of solutions gave an obstacle
when constructing second quantized theories.
As is well known this obstacle was half solved long ago in the sense
that P.~A.~M.~Dirac~\cite{dirac} resolved this problem for the case of
fermions by introducing a notion of negative energy sea, so called
``Dirac sea'',
which consists of each state completely filled by one negative energy
particle thanks to Pauli's exclusion principle.
Thus the vacuum state of fermions are identified as such a state that
all the positive energy solutions are empty and all the negative energy
ones completely filled, and thus form the Dirac sea.
Furthermore he interpreted a hole in the Dirac sea as an anti-particle.
However his Dirac sea method passing to the second quantized theory is
not applicable to boson theories simply,
because of lack of the Pauli's Principle so that infinite number of
particles can exist at each negative energy state.
Therefore one cannot define properly totally filled negative energy
states for bosons.

Needless to say that modern procedure of second quantization of fermions
as well as bosons are well established
without recourse to the Dirac sea method for fermions.
Nevertheless
to investigate a physical account of some problems in quantum field theory,
the most famous one why chiral or axial anomaly appears in fermions,
the Dirac sea method provides exceedingly well physical interpretation
of this phenomenon~\cite{Nielsen:1983rb,jackiw,'tHooft:2005cq}.
Furthermore along the line with Dirac sea method, a new peculiar effect
in quantum field theory was found. In fact the fractional fermion number 
was first discussed by R.~Jackiw~\cite{Jackiw:1984ji} and successively
Jackiw and J.~R.~Schrieffer~\cite{Jackiw:1981wc} pointed out that
this is realized in a one dimensional polymer of which properties were
investigated in detail.

It is the purpose of this article to present a consistent
formulation of the second quantization method
by introducing negative energy seas, i.e.,
Dirac seas not only for fermions but also for bosons.
In fact some of the present authors have been struggling
past few years
for obtaining consistent formulation~\cite{Nielsen:1998mc,Habara:2003cz}.
However these previous attempts had a serious drawback:
the inner product of the bosonic states of the negative energy
became indefinite.
Of course 
the indefinite norm squared 
contradicts with the principle of quantum mechanics,
and we might not be able to construct the Hilbert space.
Thus the formulation was not well unacceptable.

In the present article,
we propose a way out of the above mentioned difficulty
of the indefinite inner product.
Our new method presented in this paper provides positive definite inner
product
and the Hilbert space is consequently constructed.

We would like to stress the motivation
as to why we consider the old and
in some sense already solved problem of the method
of second quantization of relativistic quantum field theory.
The motivation is two fold:
one of them is, as is stated already above,
that the Dirac sea method provide well physical account for anomalous
phenomena due to existence of bottomless of negative energy sea,
such as chiral anomaly.
We may expect to understand as well other anomalies in case of bosons,
such as conformal anomaly.

Another motivation is that in the string theory,
in which only the light cone
gauge approach is successful in practice.
Indeed despite of many elaborate attempts,
the only light-cone string field theory
by Kaku-Kikkawa~\cite{Kaku:1974zz}
seems to be successful in the sense of consistency.
We suspect that the success of the light-cone string field theory
is due  to cutting out and disregard all negative energy modes.
We would like to apply the method presented in this article to string
theories and construct a new kind of string field theories
which will include not only positive energy modes but also negative ones.

Supersymmetry is, at first, used as a kind of device to construct each
state in boson theories as well as fermion ones at equal footing.
Therefore by imposing supersymmetry we may be able transparently to
construct the vacua of bosons like fermion's Dirac sea
which will be in fact the case
in our approach as will be described in the later sections.
Furthermore, the vacuum structure of supersymmetric theories itself is
interesting enough to investigate, i.g., whether boson and fermion vacua 
are really supersymmetric etc.

The present paper is organized as follows:
In the following Section 2,
we investigate an $N=2$ supersymmetric theory
to obtain basic relations for the boson sea.
We find that the ordinary bosonic vacuum conflicts with
the fermionic Dirac sea due to the supersymmetry.
A new condition for bosonic vacuum in terms of operator relations
are derived.
The condition indicates that
the vacuum vanishes by an operation of the creation operator
for any negative energy particle.
This means that the bosonic vacuum is just boson sea
because the condition is equivalent to that for the fermionic Dirac sea.
In Section 3,
we introduce a method of a double harmonic oscillator
to formulate the boson sea.
We extend the concept of the wave function of the harmonic oscillator
to describe not only the positive energy states
but also the negative energy ones.
The resultant vacuum corresponds with the boson sea,
which satisfies the condition derived in Section 2.
In Section 4, we consider the inner product to construct
the Hilbert space of the double harmonic oscillator.
We propose a successful definition of the positive definite inner product
by employing {\it a non-local approach}.
In Section 5,
we consider several mathematical properties of
the negative number sector.
In Section 6, 
we discuss the physical meaning of the boson sea
by comparing the Dirac sea,
and how the boson sea is filled by negative energy particles.
In Section 7,
we formulate the boson sea
as a vacuum of the second quantization of bosonic fields
by applying the double harmonic oscillator.
We confirm physical consistency of the formulation.
Section 8 is devoted to a conclusion and a brief overview of possible
future developments.

\section{Boson vacuum in a supersymmetric theory}

In this section,
we derive several basic relations for the boson sea
with recourse to supersymmetry in $4$-dimensional space-time.
We utilize the free field theory with $N=2$
supersymmetry~\cite{Sohnius:1985qm,West:1990tg},
because the simplest theory including a Dirac fermion
is $N=2$ supersymmetric one.
The fermionic vacuum is taken to be the Dirac sea.
We find difficulties of making supersymmetry
when the boson fields are quantized by the ordinary scheme
rather than our boson sea formulation.
Therefore the supersymmetry requires the boson sea.
The boson sea should be consistent with the fermionic Dirac sea
from the point of view of supersymmetry.
The consistency provides several properties of the boson sea.

Hereafter,
the Greek indices $\mu ,\nu ,\cdots$
take integer value from 0 to 3
and indicate coordinates of the Minkowski space.
Through this paper
the metric is given by $\eta^{\mu \nu}={\rm diag}(+1,-1,-1,-1)$,
and
the gamma matrices $\gamma^{\mu}$ satisfy
the ordinary anti-commutation relations.

\subsection{$N=2$ supersymmetric free theory}

Let us summarize a necessary part in this article
of a $N=2$ supersymmetric field theory in the free case.
The field contents of the theory are given
by the $N=2$ hypermultiplet~\cite{Sohnius:1985qm} as
\begin{eqnarray}
	&&(A_1, A_2;\> \psi ;\> F_1, F_2), \label{2.1}
\end{eqnarray}
where $A_i$ and $F_i$ denote complex scalar bosons with an index $i$
and $\psi$ is a Dirac fermion.
We may call $i=1,2$ the flavor index.
The multiplet (\ref{2.1}) transforms
under the following supersymmetric transformation:
\begin{eqnarray}
	\delta A_i &=& 2\bar{\xi}_i\psi,
	\nonumber\\
	\delta\psi\ &=&
	 -i\xi_iF_i-i\gamma^{\mu} \xi_i \partial_{\mu}A_i,
	\nonumber\\
	\delta F_i &=& 2\bar{\xi_i}\gamma^{\mu}\partial_{\mu}\psi,
	\label{2.2}
\end{eqnarray}
where $\xi_i$ is a Grassmann parameter of the transformation.

The Hermite self-conjugate form of the Lagrangian density
for the free theory is given by
\begin{align}
	\mathcal{L} & =\frac{1}{2}\partial_{\mu}A_i^{\dagger}\partial^{\mu}
	A_i+\frac{1}{2}F_i^{\dagger}F_i+\frac{i}{2}\bar{\psi}\gamma^{\mu}
	\partial_{\mu}\psi -\frac{i}{2}\partial_{\mu}\bar{\psi}\gamma^{\mu}
	\psi \nonumber \\
	& \qquad \qquad \qquad \qquad \qquad \qquad \quad 
	+m\left[\frac{i}{2}A_i^{\dagger}F_i-\frac{i}{2}F_i^{\dagger}A_i
	+\bar{\psi}\psi \right]. \label{Lagrangian}
\end{align}
The action of this Lagrangian density is invariant under the
supersymmetric transformation (\ref{2.2}).
The Noether current associated with the supersymmetry (\ref{2.2})
gives the supercurrent as
\begin{eqnarray}
 J_i^{\mu}
  &=&
  \gamma^{\nu}\gamma^{\mu}\psi\partial_{\nu}A_i^{\dagger}
  - i m \gamma^{\mu} \psi A_i^{\dagger},
  \label{2.11}
\end{eqnarray}
and thus the supercharges of the system are defined as
\begin{align}
	Q_i
	\equiv\int d^3\vec{x} \> J_i^0(x),
	\qquad
	\bar{Q}_i
	\equiv \int d^3\vec{x} \> \bar{J}_i^0(x),
	\label{supercharges}
\end{align}
with the flavor index $i=1,2$.
These supercharges play the important role
for deriving the property of the boson sea.

\subsection{Quantization by operator formalism}

In order to clearly specify the Dirac sea for bosons,
we briefly summarize the Dirac sea formalism for fermions
as well as
the ordinary quantization formalism for bosons.

The mode expansions of the field operators are formally given by
\begin{align}
	A_i(x)
	&=
	\int \frac{d^3\vec{k}}{\sqrt{(2\pi )^32k_0}}\left\{
	a_{i+}(\vec{k})e^{-ikx}+a_{i-}(\vec{k})e^{ikx}\right\},
	\\
	\psi(x)
	&=
	\int \frac{d^3\vec{k}}{\sqrt{(2\pi )^32k_0}}\sum_{s=\pm}
	\left\{b(\vec{k},s)u(\vec{k},s)e^{-ikx}+d(\vec{k},s)v(\vec{k},s)e^{ikx}
	\right\},
	\label{2.13}
\end{align}
where $k_0\equiv \sqrt{\vec{k}^2+m^2}$ is positive energy of the particle,
and $s\equiv {\vec{\sigma}\cdot \vec{k}}/{|\vec{k}|}$
denotes the helicity.
As is well known,
the positive energy particles are related with the operators
$a_{i+}(\vec{k})$ and $ b(\vec{k},s)$,
and the negative energy ones are related with
$a_{i-}(\vec{k})$ and $d(\vec{k},s)$.
The commutation relations among these operators are written as
\begin{eqnarray}
	\left[a_{i+}(\vec{k}),a_{j+}^{\dagger}(\vec{k}^{\prime})\right]
	&=&
	+\delta_{ij}\delta^3(\vec{k}-\vec{k}^{\prime}),
	\label{B+com}\\
	\left[a_{i-}(\vec{k}),a_{j-}^{\dagger}(\vec{k}^{\prime})\right]
	&=&
	-\delta_{ij}\delta^3(\vec{k}-\vec{k}^{\prime}),
	\label{B-com}\\
	\left\{b(\vec{k},s),b^{\dagger}(\vec{k}^{\prime},s^{\prime})\right\}
	\ &=&
	+\delta_{ss^{\prime}}\delta^3(\vec{k}-\vec{k}^{\prime}),
	\label{F+com}\\
	\left\{d(\vec{k},s),d^{\dagger}(\vec{k}^{\prime},s^{\prime})\right\}
	\: &=&
	+\delta_{ss^{\prime}}\delta^3(\vec{k}-\vec{k}^{\prime}),
 	\label{F-com}\label{2.17}
\end{eqnarray}
while all other pairs are commuting or anti-commuting.
It is important 
that the right-hand side of the commutation relation (\ref{B-com})
for negative energy bosons
has the opposite sign to those of (\ref{B+com}) for positive energy bosons.

In the ordinary context,
these operators are naively interpreted as follows:
\begin{align*}
	& \left\{ \begin{array}{l}
	a_{i+}(\vec{k})\text{ annihilates a bosonic particle 
	with energy-momentum $(k_0,\vec{k})$,} \\
	a_{i+}^{\dagger}(\vec{k})\text{ creates a bosonic particle 
	with energy-momentum $(k_0,\vec{k})$,} \\
	a_{i-}(\vec{k})\text{ annihilates a bosonic particle 
	with energy-momentum $(-k_0,-\vec{k})$,} \\
	a_{i-}^{\dagger}(\vec{k})\text{ creates a bosonic particle 
	with energy-momentum $(-k_0,-\vec{k})$,} \\
	\end{array} \right. \\
	& \left\{ \begin{array}{l}
	b(\vec{k},s)\text{ annihilates a fermionic particle 
	with energy-momentum $(k_0,\vec{k})$,} \\
	b^{\dagger}(\vec{k},s)\text{ creates a fermionic particle 
	with energy-momentum $(k_0,\vec{k})$,} \\
	d(\vec{k},s)\text{ annihilates a fermionic particle 
	with energy-momentum $(-k_0,-\vec{k})$,} \\
	d^{\dagger}(\vec{k},s)\text{ creates a fermionic particle 
	with energy-momentum $(-k_0,-\vec{k})$.}
	\end{array} \right.
\end{align*}
Here the particles with the energy-momentum $(-k_0,-\vec{k})$
have negative energy $(-k_0)$,
because $k_0=\sqrt{\vec{k}^2+m^2}$ is defined as always positive one.

The total vacuum $||0\rangle$ of the system is decomposed into\footnotemark
\begin{eqnarray}
	||0\rangle
	&\;\equiv\;&
	||0_{+}\rangle \otimes
	||0_{-}\rangle \otimes 
	||\tilde{0}_{+}\rangle \otimes
	||\tilde{0}_{-}\rangle, 
	\label{2.23}
\end{eqnarray}
where $\otimes$ denotes the tensor product, and
the notation of the sub-vacua is the following:
\begin{eqnarray*}
 &
 \begin{array}{ll}
 ||0_{\pm}\rangle &
  \mbox{: the boson vacua for positive and negative energy states
  respectively},\\
 ||\tilde{0}_{\pm}\rangle &
  \mbox{: the fermion vacua for positive and negative energy states
  respectively}.
 \end{array}
 &
\end{eqnarray*}
\footnotetext{%
In the following, for example in the boson case,
we denote the vacua by
$|0_{\pm}\rangle$ in the system of single particle, and
$||0_{\pm}\rangle$ in the system with many particles.}%

Here, we adopt formulation of the Dirac sea for fermions as follows:
needless to say,
the fermion vacuum consists of
the positive energy part $||\tilde{0}_{+}\rangle$ and
the negative energy part $||\tilde{0}_{-}\rangle$.
The positive energy vacuum $||\tilde{0}_{+}\rangle$
is an empty state which is defined as
\begin{eqnarray}
 b(\vec{k},s) ||\tilde{0}_{+}\rangle &=& 0.
 \label{fermion_vacuum_condition+}
\end{eqnarray}
The negative energy vacuum $||\tilde{0}_{-}\rangle$
is nothing but the Dirac sea
and it is constructed by acting all $d^{\dagger}$'s
on the empty state $||\widetilde{\emp}\rangle$
to give the expression
\begin{align}
	||\tilde{0}_{-}\rangle
	\equiv
	\bigg\{ \prod_{\vec{k},s} d^{\dagger}(\vec{k},s) \bigg\}
 	||\widetilde{\emp}\rangle.
	\label{2.23.5}
\end{align}
The empty state used in (\ref{2.23.5})
is defined by $d(\vec{p},s)||\widetilde{\emp}\rangle = 0$.
The Dirac sea $||\tilde{0}_{-}\rangle$ satisfies
\begin{eqnarray}
 d^{\dagger}(\vec{p},s) ||\tilde{0}_{-}\rangle &=& 0.
 \label{fermion_vacuum_condition-}
\end{eqnarray}
We have used $d^{\dagger}$ as the creation operator
and $d$ as the annihilation operator for negative energy fermions.
It is of importance to remind that
the operators $d^{\dagger}$ and $d$ are re-interpreted as
the annihilation operator and creation operator respectively
for holes with positive energy.

We briefly review the ordinary quantization scheme for bosons,
for the sake of the contrast to our new method
which will be shown in the next subsection.
We introduce new operators
\begin{align}
	{\alpha}_{i-}           \equiv a_{i-}^{\dagger}, \qquad
	{\alpha}_{i-}^{\dagger} \equiv a_{i-}
\end{align}
for the negative energy particles.
This introduction of these operators is essential
in the scheme,
because
the role of the creation and annihilation operators is inverted.
The commutation relation (\ref{B-com}) is rewritten into
\begin{align}
	\left[{\alpha}_{i-}(\vec{k}),{\alpha}_{j-}^{\dagger}
	(\vec{k}^{\prime})\right]=+\delta_{ij}\delta^3
	(\vec{k}-\vec{k}^{\prime}).
	\label{2.19}
\end{align}
The introduction of the operators
${\alpha}_{i-}$ and ${\alpha}_{i-}^{\dagger}$
allows us to treat negative energy bosons
in the same manner as positive energy bosons,
because
the right-hand side of the commutator (\ref{2.19}) has positive sign.
The vacuum for positive and negative energy bosons,
which are denoted
$||0_{+}^{\text{ordinary}}\rangle$ and
$||0_{-}^{\text{ordinary}}\rangle$ respectively,
are defined by
\begin{align}
	& a_{i+} (\vec{k}) ||0_{+}^{\text{ordinary}}\rangle=0,
	\label{2.17.5}\\
	& {\alpha}_{i-} (\vec{k}) ||0_{-}^{\text{ordinary}}\rangle=0. 
	\label{2.18.5}
\end{align}
In this ordinary scheme,
both
the positive energy vacuum $||0_{+}^{\text{ordinary}}\rangle$ and
the negative energy vacuum $||0_{-}^{\text{ordinary}}\rangle$
are empty states.

\subsection{Supersymmetry and a condition of boson sea}

As described in the previous subsection,
the vacuum of bosons in the ordinary quantization scheme
is quite different from the Dirac sea as the vacuum of fermions.
In this subsection,
we find a conflict between these pictures
when we require the supersymmetry.
The supersymmetry requires the boson sea,
and determines properties of it.

In terms of the creation and annihilation operators,
the supercharges given by (\ref{supercharges}) become
\begin{align}
	Q_i
	& =+i\int d^3\vec{k} \sum_{s=\pm} 
	\left\{b(\vec{k},s)u(\vec{k},s)
	a_{i+}^{\dagger}(\vec{k})-d(\vec{k},s)v(\vec{k},s)a_{i-}^{\dagger}
	(\vec{k})\right\}, \nonumber\\
	\bar{Q}_i
	& =-i\int d^3\vec{k} \sum_{s=\pm} 
	\left\{b^{\dagger}(\vec{k},s)
	\bar{u}(\vec{k},s)a_{i+}(\vec{k})-d^{\dagger}(\vec{k},s)\bar{v}
	(\vec{k},s)a_{i-}(\vec{k})\right\}.
	\label{2.21}
\end{align}
The condition for the vacuum to be supersymmetric is
\begin{align}
	Q_i||0\rangle=\bar{Q}_i||0\rangle=0. \label{2.22}
\end{align}
By combining this condition (\ref{2.22}) with
the properties (\ref{fermion_vacuum_condition+})
and (\ref{fermion_vacuum_condition-}) of the Dirac sea,
the condition of the supersymmetric vacuum for the boson reads
\begin{align}
	& a_{i+}(\vec{k})||0_{+}\rangle=0,
	\label{2.24}\\
	& a_{i-}^{\dagger}(\vec{k})||0_{-}\rangle=0.
	\label{2.25}
\end{align}
The vacuum condition (\ref{2.25})
are apparently \underline{inconsistent}
with the vacuum condition (\ref{2.18.5}) for the ordinary boson scheme.
In fact, in ordinary theory as one can see,
the negative energy vacuum $||0_{-}^{\text{ordinary}}\rangle$ is annihilated
by annihilation operator ${\alpha}_{i-}(\vec{k})$,
while in our method
the corresponding vacuum $||0_{-}\rangle$
is annihilated by the creation operator $a_{i-}^{\dagger}(\vec{k})$.
Therefore,
we just find that
the ordinary boson scheme 
conflicts with the Dirac sea formulation
that respects the supersymmetry.

The conditions (\ref{2.24}) and (\ref{2.25})
seems to be an evidence for boson sea.
These conditions are the same as 
the relations
(\ref{fermion_vacuum_condition+}) and
(\ref{fermion_vacuum_condition-}) for the Dirac sea.
The vacuum $||0_{+}\rangle$ corresponds the empty vacuum
which vanishes by the annihilation operator $a_{i+}$.
The vacuum $||0_{-}\rangle$ vanishes by $a_{i-}^{\dagger}$
which creates the negative energy quantum.

The condition (\ref{2.25}) is a result of the supersymmetry.
We may well generalize it,
and consider that the condition (\ref{2.25}) is
a general condition of the boson sea $||0_-\rangle$.
We call the usual systems
characterized by the algebra (\ref{B+com}) and the condition (\ref{2.24})
the ``positive number sector'' of the positive energy particle states,
while
we call the unusual system characterized by 
the algebra (\ref{B-com}) and the condition (\ref{2.25})
the ``negative number sector'' of the negative ones.
In the next section,
we explicitly construct the negative number sector.

\section{Double harmonic oscillator}

In this section, we construct the negative number sector
which is required in the boson sea formulation.
The sector is obtained by extending
the harmonic oscillator spectrum to negative energy\footnotemark.
In particular,
an extension of the wave function plays an important role,
and
the resultant sector leads to the boson sea.
\footnotetext{%
See Ref.~\cite{vanHolten:1984ia} for another negative energy solution of
(deformed) harmonic oscillator.}

\subsection{Analytic wave function of the harmonic oscillator for
negative energy}

Although most of the contents were already described
in Ref.~\cite{Nielsen:1983rb,Nielsen:1998mc},
we review those for self-containedness.
The Schr\"{o}dinger equation of the one-dimensional harmonic oscillator
is
\begin{align}
	\left(
	-\frac{1}{2}\frac{\partial^2}{\partial x^2}+\frac{1}{2}x^2
	\right) 
	\phi (x)=E\phi (x).
	\label{3.2}
\end{align}
The ordinary solutions are given by
$A_n H_n(x)e^{-\frac{1}{2}x^2}$,
where $H_n(x)$ denotes the Hermite polynomial
and $A_n$ is the normalization factor.
In order to find further solutions, we assume the form 
\begin{align}
	\phi (x)=f(x)e^{\pm \frac{1}{2}x^2}.
	\label{3.3}
\end{align}
%
%
By substituting (\ref{3.3}) into (\ref{3.2}),
we obtain the following equation for $f(x)$: 
\begin{align}
	\frac{f^{\prime\prime}(x)}{f(x)}
	\pm
	\frac{2xf^{\prime}(x)}{f(x)}
	=
	-2 \left(E\pm \frac{1}{2}\right).
	\label{3.4}
\end{align}

We derive asymptotic behavior of $f(x)$ for large $|x|$.
We assume that the second term on the left-hand side
dominates the first term, when $|x|$ is large.
The equation (\ref{3.4}) becomes
\begin{align}
	\frac{d\log f(x)}{d\log x}
	=
	\mp E-\frac{1}{2},
	\label{3.5}
\end{align}
and we find the asymptotic behavior 
as
\begin{align}
	f(x)\sim x^{\left(\mp E- \frac{1}{2}\right)}.
	\label{3.7}
\end{align}

We consider an analytic continuation of $x$ to the whole complex plane,
and
require $f(x)$ to be a single-valued analytic function
all over the complex plane.
This analytic continuation restricts the power in (\ref{3.7}),
which means that
$(\mp E- \frac{1}{2})$ should be zero or positive integer.
We find the quantized energy
\begin{align}
	E
	=
	\mp \left( n + \frac{1}{2} \right) ,
	\qquad
	n=0,1,2,\cdots.
	\label{3.8}
\end{align}
This energy spectrum includes not only the positive part
but also the negative one.

As is well known,
the ordinary harmonic oscillator is characterized
by the positive energy spectrum
\begin{align}
	E = n + \frac{1}{2},\qquad n=0,1,2,\cdots,
\end{align}
which corresponds with (\ref{3.8}) with the lower sign.
The eigenfunctions are 
\begin{align}
	\phi_n(x)
	=
	A_n
	H_n(x)
	e^{-\frac{1}{2}x^2},
	\label{3.10}
\end{align}
where $A_n$ is the normalization factor%
\footnote{
The normalization factor for the ordinary harmonic oscillator becomes
$A_n = \left( \sqrt{\pi}2^nn!\right)^{-1/2}$,
however, we adopt different definition of the norm
due to the consistency with the negative number sector.}.
We call this positive energy spectrum {\it the positive number sector}.

\begin{figure}[ht]
	\begin{center}
	\includegraphics{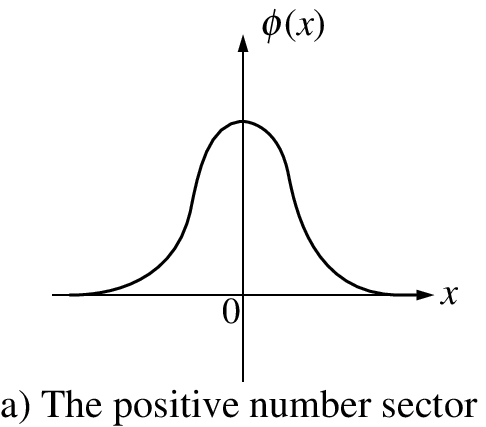}
	\hspace{3em}
	\includegraphics{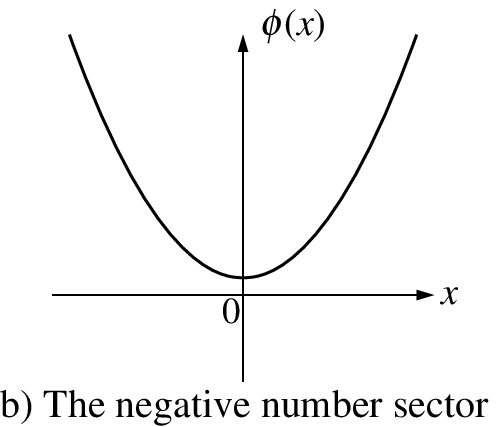}
	\end{center}
	\caption{Wave functions of the ground states (the vacua).}
	\label{gauss}
\end{figure}

In the following,
we consider the negative energy spectrum or
{\it the negative number sector}.
The negative energy spectrum is described by (\ref{3.8})
with the upper sign as
$E = -\left( n+\frac{1}{2}\right)$,
and the eigenfunction is given by (\ref{3.3}) with the upper sign as
$\phi (x) = f(x)e^{+\frac{1}{2}x^2}$.
We determine the explicit form of $f(x)$ in the following:
The negative number sector is formally given
by simple replacement $x \to ix$
in the positive number sector:
\begin{align}
	&
	\left(
		-\frac{1}{2}\frac{\partial^2}{\partial(ix)^2}
		+\frac{1}{2}(ix)^2
	\right) 
	\phi (ix) = E\phi (ix)
	\ \ \Longrightarrow\ \ 
	\left(
		-\frac{1}{2}\frac{\partial^2}{\partial x^2}
		+\frac{1}{2}x^2
	\right)
	\phi (ix) = -E\phi (ix).
	\label{3.12}
\end{align}
By this replacement,
the eigenfunctions and eigenvalues of the negative number sector
are obtained as
\begin{eqnarray}
	&&
	E
	= -\left( n+\frac{1}{2}\right),\qquad n=0,1,2,\cdots,
	\nonumber\\
	&&
	\phi_{-n}(x)
	=
	A_{-n} H_n(ix) e^{-\frac{1}{2}(ix)^2}.
	\label{3.14}
\end{eqnarray}
The Hermite polynomial $H_n(ix)$ in (\ref{3.14}) is
either purely real or purely imaginary,
because $H_n(x)$ is either an even or odd function of $x$.
By extending the Hermite polynomial to negative $n$ as
\begin{eqnarray}
 H_{n}(x)
  &=&
  \left\{
   \begin{array}{ll}
    H_n(x)          & (n \geq 0)\\
    i^n H_{-n}(i x) & (n < 0)
   \end{array}
  \right.,
  \label{HermitePolynomial}
\end{eqnarray}
the extended Hermite polynomial becomes purely real function.
By using this polynomial
and by redefining the normalization factor $A_{-n}$,
we rewrite the eigenfunctions of the negative number sector (\ref{3.14}) 
in terms of real functions
\begin{eqnarray}
 \phi_{-n}(x)
  &=& A_{-n} H_{-n}(x) e^{+\frac{1}{2}x^2}, \label{3.14new}
\end{eqnarray}
The wave functions of the ground states are depicted in Fig.~\ref{gauss}
for the positive and the negative number sectors.

In the usual quantum mechanics,
the wave function $\phi(x)$ should be normalized as
square integrable
\begin{align}
	\int_{-\infty}^{+\infty}dx \> |\phi (x)|^2 < +\infty.
	\label{3.1}
\end{align}
The eigenfunctions in the positive number sector satisfy
this normalization condition.
However,
those of the negative number sector do not satisfy the condition
(\ref{3.1}),
due to the positive exponent $e^{+x^2}$
of the eigenfunctions (\ref{3.14new}).
When this condition is imposed,
the one-dimensional harmonic oscillator turns out to become
only positive number sector.
Therefore we should extend the condition (\ref{3.1})
in order to obtain the negative number sector.

\subsection{Representation of the double harmonic oscillator}

We formulate a method of the double harmonic oscillator,
which allows us to treat
both positive and negative number sectors of a harmonic oscillator.
The total Hilbert space is constructed by operator formalism.

The negative number sector is independent of the positive one,
because any state in the negative number sector cannot be derived
by finite operations of creation and annihilation operators
on the positive number sector.
Therefore it is natural to introduce
{\it the double harmonic oscillators}
which consist of the two independent harmonic oscillators.
The system is described by the two-dimensional harmonic oscillator.
The two-dimensional space of the oscillator has
an Minkowski-like indefinite metric whose elements are given by
$\eta_{xx}=1$, $\eta_{yy}=-1$ and $\eta_{xy}=\eta_{yx}=0$.
One of the harmonic oscillators describes the positive number sector
and the other is negative number sector.

To implement of this idea
we introduce an algebra with an indefinite metric.
The algebra consists of two elements $I$ and $J$,
which satisfy the following relations:
\begin{eqnarray}
 I^2 = I,\qquad J^2 = -I,\qquad J I = I J = J.
 \label{algebra1}
\end{eqnarray}
The indefinite metric is defined as
\begin{eqnarray}
 &&\left<I,I\right> =  1,\qquad
   \left<J,J\right> = -1,\qquad
   \left<I,J\right> = \left<J,I\right> = 0,
 \label{algebra2}
\end{eqnarray}
where $\left<,\right>$ denotes the metric in the form of a product.
This metric is an ordinary bilinear product and has the property
\begin{eqnarray}
  \left< (I \otimes a + J \otimes b), (I \otimes c + J \otimes d) \right>
   &=&
  a^* c - b^* d.
\end{eqnarray}
with the tensor product $\otimes$.
By combining (\ref{algebra1}) and (\ref{algebra2}),
we find that the elements are self-adjoint
\begin{eqnarray}
 I^\dagger = I, \qquad  J^\dagger = J,
\end{eqnarray}
because of
$1
= \left<I,I\right>
= -\left<I,JJ\right>
= -\left<J^\dagger I,J\right>
= -\left<J^\dagger,J\right>$.
The element $I$ is identified as identity, while
the element $J$ can be identified as the imaginary unit
with negative norm squared.
The Hilbert space and the operator space
are extended by this algebra.

The system of the double harmonic oscillator is
constructed on the two dimensional space spanned by
\begin{eqnarray}
 z &\equiv& I \otimes x + J \otimes y,
  \label{Minkowski}
\end{eqnarray}
where the coordinates $x$ and $y$ are real numbers.
This is just the two-dimensional Minkowski space.
The Laplacian on this space is defined as
$\triangle_z \equiv \partial_x^2 - \partial_y^2$,
and the product is $z \cdot z = x^2 - y^2$.
The Schr\"{o}dinger equation becomes
\begin{align}
	 I \otimes
	\left\{
	 \left( -\frac{1}{2}\frac{\partial^2}{\partial x^2}
		+\frac{1}{2}x^2 \right)
	 \:-\:
	 \left( -\frac{1}{2}\frac{\partial^2}{\partial y^2}
		+\frac{1}{2}y^2 \right)
	\right\}
	\phi (x,y)
	\ =\ 
	I \otimes
	E\phi (x,y).
	\label{3.31}
\end{align}
The coordinates $x$ and $y$ describe
the positive and negative number sectors respectively.

It is important that
any state of the negative number sector also has positive energy.
The negative number excitation of the negative energy state
has positive energy,
because the negative integer times the negative energy becomes positive.
This important property is realized by
the combination of the negative number sector (\ref{3.14new})
and the Minkowski space (\ref{Minkowski}).
This property reproduces
that anti-particles as holes in the boson sea have positive energy.

\begin{figure}[htbp]
 \begin{center}
  \includegraphics{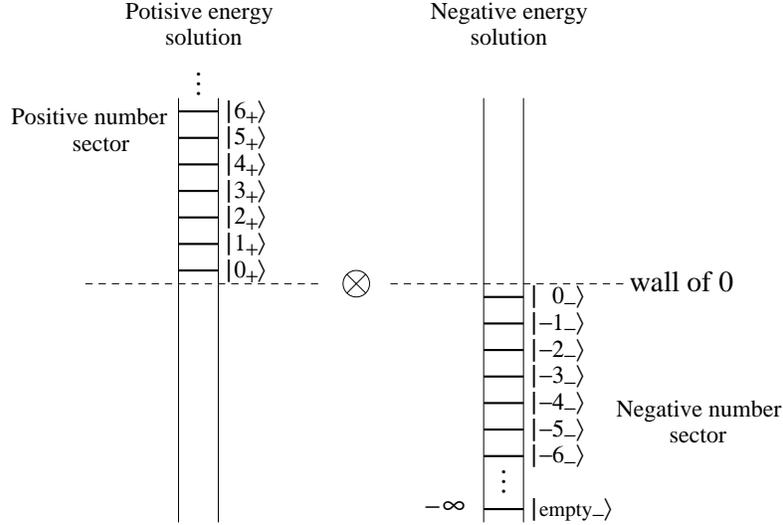}
 \end{center}
 \caption{%
 All states of the double harmonic oscillator.
 The state $\left|n_{+},-m_{-}\right>$ is formally denoted as
 the tensor product
 $\left|n_{+}\right> \otimes \left|-m_{-}\right>$.
 The vertical axis means the number of the excitations (the quanta).
 All of the state have positive energy (\ref{TotalEnergyValue}).
 }
 \label{tower}
\end{figure}

We denote a state $\left|n_{+},-m_{-}\right>$
as a number-eigenstate
which consists of $n$ quanta in the positive number sector
and $(-m)$ quanta in the negative number sector,
where $n,m = 0,1,2,\cdots$.
The subscript $\pm$ indicates
whether the number belongs to the positive number sector or
the negative one.
The corresponding wave function of the state is denoted as
$\phi_{n_+,-m_-}$.
The positive and negative number states have eigenenergy
$E_{n_+} = \frac12 + n$ and $E_{-m_-} = -\frac12 -m$ respectively.
The eigenenergy of the state becomes
\begin{eqnarray}
 E_{n_+,-m_-} &=& E_{n_+} + E_{-m_-} \times (-1) \;=\; 1 + n + m,
  \label{TotalEnergyValue}
\end{eqnarray}
where the factor $(-1)$ in the second term
comes from the indefinite metric.
All of the states are shown in Fig.~\ref{tower}.

The vacuum of the system is given by
\begin{eqnarray}
 \phi_{0_+,0_-}(x,y)
  &\equiv&
  I \otimes \: 
  e^{-\frac{1}{2}x^2+\frac{1}{2}y^2},
  \label{vacuum_state}
\end{eqnarray}
whose eigenvalue of energy is $E_{0_+,0_-} = 1$.
This corresponds with the vacuum energy of two bosons system
which consists of a boson and an anti-boson.

The annihilation and creation operators of the positive number sector
are 
\begin{eqnarray}
 a_+
 &=& I \otimes \frac{1}{\sqrt{2}}\left(x + \frac{\partial}{\partial x}\right),
 \qquad
 a_+^{\dagger}
  =  I \otimes \frac{1}{\sqrt{2}}\left(x - \frac{\partial}{\partial x}\right)
\end{eqnarray}
respectively, which satisfy the ordinary commutation relation
\begin{eqnarray}
 \left[a_+, \; a_+^\dagger\right] &=& +I.
\end{eqnarray}
On the other hand, 
the annihilation and creation operators of the negative number sector
are given by
\begin{eqnarray}
 a_-
  &=& J \otimes \frac{1}{\sqrt{2}}\left(y + \frac{\partial}{\partial y}\right),
 \qquad
 a_-^{\dagger}
   =  J \otimes \frac{1}{\sqrt{2}}\left(y - \frac{\partial}{\partial y}\right).
 \label{a_adagger_negative}
\end{eqnarray}
These operators satisfy
\begin{eqnarray}
 \left[a_-, \; a_-^\dagger\right] &=& -I,
 \label{com_a_a_negative}
\end{eqnarray}
which just reproduces the relation (\ref{B-com})
for negative energy solutions.

We confirm that the operators $a_-$ and $a_-^\dagger$
of the negative number sector work properly in the following:
The vacuum is annihilated by the creation operator as
\begin{eqnarray}
	a_-^{\dagger}\phi_{0_+,0_-}(x,y)
	&=& J \otimes \frac{1}{\sqrt{2}} 
		\left(y - \frac{\partial}{\partial y}\right)
		e^{-\frac{1}{2}x^2+\frac{1}{2}y^2}
	 =  0,
	 \label{negative_vacuum_vanish}
\end{eqnarray}
and the annihilation operator $a_-$
creates the negative excited states as
\begin{eqnarray}
	\phi_{0_+,-m_-}(x,y)
	 &=&
	 \frac{1}{\sqrt{m!}}
	 (a_-)^m \phi _{0_+,0_-}(x,y) \nonumber\\
	 &=&
	 J^m \otimes \frac{1}{\sqrt{2^{m}m!}} 
	 H_{-m}(y) \:
	 e^{-\frac{1}{2}x^2+\frac{1}{2}y^2}
	  \label{3.18}
\end{eqnarray}
for the integer $m \geq 1$.
We can also confirm that
the creation operator $a_-^\dagger$
annihilates the negative excited sates:
\begin{eqnarray}
 \phi_{0_+,(-m+1)_-}(x,y)
  &\propto& a_-^{\dagger}\phi_{0_+,-m_-}(x,y),
  \label{3.19}
\end{eqnarray}
Therefore we find that
$a_-$ creates a negative energy quantum and
$a_-^{\dagger}$ annihilates a negative energy quantum.

The total space of the double harmonic oscillator
is constructed on the vacuum state (\ref{vacuum_state}) by
the creation and annihilation operators
as is depicted in the diagram in Fig.~\ref{FockTree}.
The positive number sector exactly corresponds
with the quantum mechanics of the ordinary harmonic oscillator.
The operator $a_+$ ($a_+^{\dagger}$) annihilates (creates)
a quantum of the positive energy.
On the other hand,
the operator $a_-$ ($a_-^{\dagger}$) for negative number sector
creates (annihilates) the negative energy quantum.
The general form of the wave function is given by
\begin{eqnarray}
	\phi_{n_+,-m_-}(x,y)
	 &=&
	 \frac{1}{\sqrt{n!m!}}
	 (a_+^\dagger)^n (a_-)^m
	 \phi _{0_+,0_-}(x,y) \nonumber\\
	 &=&
	 J^m \otimes \frac{1}{\sqrt{2^n 2^m n! m!}} 
	 H_{n}(x) H_{-m}(y) \:
	 e^{-\frac{1}{2}x^2+\frac{1}{2}y^2}.
	 \label{general-wavefunc}
\end{eqnarray}

\newcommand{\LONGUD}{%
\scriptstyle a_-%
\rotatebox[origin=c]{90}{$\longrightarrow$}
\rotatebox[origin=c]{90}{$\longleftarrow$}
\scriptstyle a_-^{\dagger}%
}%
\newcommand{\LONGD}{%
\scriptstyle\rule{1.6em}{0em}%
\rotatebox[origin=c]{90}{$\longleftarrow$}%
\scriptstyle a_-^{\dagger}%
}%
\newcommand{\LONGLR}{%
\begin{array}{c}%
 \scriptstyle a_+^{\dagger}\\[-0.5em]%
 \longrightarrow\\[-0.8em]%
 \longleftarrow\\[-0.8em]%
 \scriptstyle a_+%
\end{array}%
}%
\newcommand{\LONGL}{%
\begin{array}{@{}c}%
 \rule{0em}{0.5em}\\[-0.5em]%
 \rule{1.4em}{0em}\\[-0.8em]%
 \longleftarrow\\[-0.8em]%
 \scriptstyle a_+%
\end{array}%
}%
\begin{figure}[htbp]
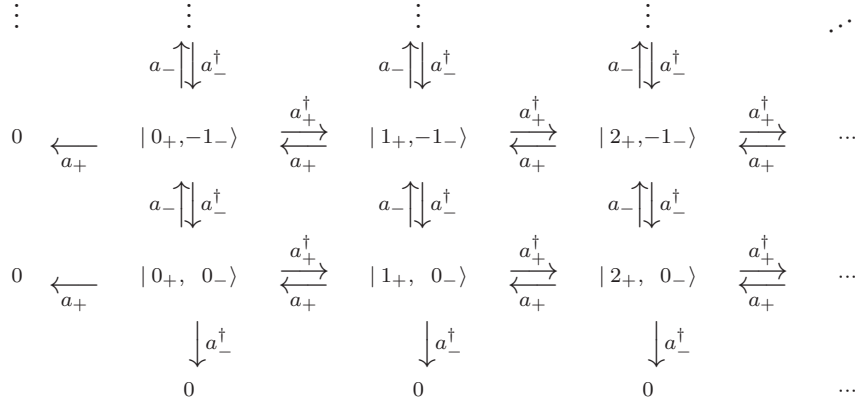

\begin{center}
 $
  \begin{array}{ccccccccc}
   \scriptstyle\vdots
    &&
    \scriptstyle\vdots
    &&
    \scriptstyle\vdots
    &&
    \scriptstyle\vdots
    &&
    \scriptstyle\rotatebox{-60}{\vdots}
    \\
   && \LONGUD && \LONGUD && \LONGUD && 
    \\
   \scriptstyle 0 & \LONGL &
    \scriptstyle\left|\:0_+,-1_-\right> &
    \LONGLR &
    \scriptstyle\left|\:1_+,-1_-\right> &
    \LONGLR &
    \scriptstyle\left|\:2_+,-1_-\right> &
    \LONGLR &
    \scriptstyle\cdots
    \\
   && \LONGUD && \LONGUD && \LONGUD && 
    \\
   \scriptstyle 0 & \LONGL &
    \scriptstyle\left|\:0_+,\ \; 0_-\right> &
    \LONGLR &
    \scriptstyle\left|\:1_+,\ \; 0_-\right> &
    \LONGLR &
    \scriptstyle\left|\:2_+,\ \; 0_-\right> &
    \LONGLR &
    \scriptstyle\cdots
    \\
   && \LONGD && \LONGD && \LONGD && 
    \\
   && \scriptstyle0 && \scriptstyle0 && \scriptstyle0 && \scriptstyle\cdots
  \end{array}
$
\end{center}
 \caption{Construction of the total space of the double harmonic oscillator.}
 \label{FockTree}
\end{figure}

\section{Inner product of the System}

Let us define the inner product 
to construct the Hilbert space of the double harmonic oscillator.
A naive definition of the inner product for wave functions $f$ and $g$ 
would be
\begin{eqnarray}
 \langle f \:|\: g \rangle 
  &=&
  \int_{-\infty}^{+\infty}dx \int_{-\infty}^{+\infty}dy \> 
  \left<
   f (x,y),\:
   g (x,y)
  \right>,
   \label{naive-product}
\end{eqnarray}
however,
this product diverges because of the factor $e^{+\frac{1}{2}y^2}$ 
in the negative number sector.
Thus we should regularize the product (\ref{naive-product}).
The product (\ref{naive-product}) has another problem:
The product (\ref{naive-product}) is not positive definite,
due to the negative norm of $J$-element, namely, $\left<J,J\right>=-1$.

The consistency among
the commutation relation (\ref{com_a_a_negative})
and the vacuum property (\ref{negative_vacuum_vanish})
for the negative number sector
requires the positive definite inner product.
For example,
the norm of $|0_+,-1_-\rangle$
should correspond with that of the vacuum as
\begin{eqnarray}
 \langle 0_+,-1_- | 0_+,-1_-\rangle 
 &=& \langle 0_+,0_-|a_-^{\dagger}a_-|0_+,0_-\rangle
 \;=\; \langle 0_+,0_-|(a_-a_-^{\dagger}+I)|0_+,0_-\rangle\nonumber\\
 &=& \langle 0_+,0_-|0_+,0_-\rangle.
\end{eqnarray}
Therefore 
the naive definition ({naive-product}) of the inner product
is ruled out.

The inner product
proposed in Ref.~\cite{Nielsen:1998mc,Habara:2003cz,nn3}
is another candidate.
In the definition,
the wave function is analytically continued 
into the whole complex plane,
and the integration is chosen along the pure imaginary axis:
\begin{eqnarray}
	\langle f\:|\:g \rangle
	&\propto&
	\int_{-\infty}^{+\infty}dx\> 
	\int_{-i\infty}^{+i\infty}dy\> 
	f^{*}(x,y) g(x,y) 
	\label{3.30}
\end{eqnarray}
The product (\ref{3.30}) gives finite value,
because the factor $e^{+\frac{1}{2}y^2}$
becomes the Gaussian type $e^{-\frac{1}{2}y^2}$
on the pure imaginary axis.
However,
the product (\ref{3.30}) is not positive definite,
and the negative number sector is treated
as the positive number sector.
Therefore
this treatment can not provide a physical picture of the boson sea 
and is regrettably not suitable for our approach.

In the following subsections,
we propose a successful definitions of the inner product:
We employ {\it a non-local approach}.
The definition of the inner product allows us to
construct the Hilbert space,
because it provides a positive definite inner product.

\subsection{Inner product by Non-local Approach}

In this subsection we present a definition of
the positive definite inner product
by employing a non-local method.
We begin by defining the inner product as
\begin{eqnarray}
 \left< f \:|\: g \right>
 &=&
 \int_{-\infty}^{+\infty} dx
 \int_{-\infty}^{+\infty} dy_1
 \int_{-\infty}^{+\infty} dy_2
 \: G(y_1, y_2)
 \left< f(x,y_1), \: g(x,y_2) \right>,
 \label{G-product}
\end{eqnarray}
where we have introduced the integral kernel $G(y_1,y_2)$
which has real value and is symmetric function: $G(y_1,y_2) = G(y_2,y_1)$.
We allow the non-local contribution of $y$-coordinate
in this definition.
The ordinary inner product is reproduced
by taking $G(y_1,y_2) = \delta(y_1 - y_2)$.

The integral kernel $G(y_1,y_2)$ is regarded as a metric tensor
for the negative number sector,
whose indices are $y_1$ and $y_2$.
All of the properties of the inner product
for the negative number sector
is governed by the metric tensor $G(y_1,y_2)$.
The metric tensor $G(y_1,y_2)$ is determined
so that
the inner product (\ref{G-product})
should satisfy the ortho-normal condition:
\begin{eqnarray}
 \left< n_+, -m_- | n_+', -m_-' \right>
  &=&  \delta_{n,n'} \delta_{m,m'}.
  \label{orthonormal-condition}
\end{eqnarray}
Therefore $G(y_1,y_2)$ plays
the role of a regularization of the divergence in (\ref{naive-product})
and makes the inner product (\ref{G-product}) be positive definite.

We are going to present how to determine the metric tensor $G(y_1,y_2)$
in the following part of this subsection.
%
%
We take the metric tensor
\begin{eqnarray}
 G(y_1,y_2) &=& \Lambda(y_1) \Lambda(y_2) \: g(y_1,y_2),
\end{eqnarray}
where the function
\begin{eqnarray}
 \Lambda(y) &\equiv& e^{-y^2}
\end{eqnarray}
is a regularization factor,
and the real symmetric function $g(y_1,y_2)$ is the principal part of
the metric tensor.
To realize the ortho-normal condition (\ref{orthonormal-condition}),
$g(y_1,y_2)$ should satisfy the following condition:
\begin{eqnarray}
 \int dy_1 dy_2 \; f_{n}(y_1) g(y_1,y_2) f_{m}(y_1) &=& {\cal I}_{nm}
  \label{f-product}
\end{eqnarray}
for $n,m=0,1,2,\cdots$.
Here we have temporarily defined the regularized wave function
of the negative number sector as
\begin{eqnarray}
 f_n(y)
  &\equiv&
  \frac{1}{\pi^{1/4}\sqrt{2^{n}\:n!}} e^{-\frac{1}{2}y^2} H_{-n}(y)
  \label{f_n}
\end{eqnarray}
for $n=0,1,2,\cdots$,
and we have also defined the unit matrix with the parity correction:
\begin{eqnarray}
 {\cal I}_{nm} &=& \delta_{nm} (-1)^n.
\end{eqnarray}
This assignment of ${\cal I}_{nm}$ makes the product (\ref{G-product})
be positive definite,
because
the negative sign from the product $\left<J,J\right>$
arises when $n$ is odd.

To utilize technique of the linear algebra,
we denote $y$ as a subscript,
and we take the contraction rule for the subscript.
Thus the equation (\ref{f-product}) is simply rewritten as
\begin{eqnarray}
 f_{ny_1} g_{y_1 y_2} f_{my_2} &=& {\cal I}_{nm}.
  \label{f-product-2}
\end{eqnarray}
While the index $y$ of the matrix description $f_{ny}$
of the function $f_{n}(y)$ 
is continuous variable rather than integer,
the matrix $f_{ny}$ is regarded as a square matrix of infinite dimensions.
The regularized wave functions $\{f(y)\}$ belong to the Hilbert space
because a regularized wave function $f(y)$ is square-integrable,
and the zero-norm functions are identified with the zero vector of
the Hilbert space.
The degree of the freedom of $\{f(y)\}$ is equals to that of
$\{f_n\}$, thus $f_{ny}$ is square matrix in this sense.

By applying the inverse matrix of $f$ to the equation (\ref{f-product-2}),
the metric $g_{y_1 y_2}$ is obtained as
\begin{eqnarray}
 g_{y_1 y_2} &=& (f^{-1})_{y_1n} {\cal I}_{nm} (f^{-1})_{y_2m},
  \label{g-by-f-inverse}
\end{eqnarray}
where the inverse matrix $f^{-1}$ satisfies
\begin{eqnarray}
 (f^{-1})_{y_1n} \: f_{ny_2} &=& 
  \delta(y_1-y_2),\\
 f_{ny} \: (f^{-1})_{ym} &=& \delta_{nm}.
\end{eqnarray}

We define the ordinary inner product of $f_n$ and $f_m$ as
\begin{eqnarray}
 P_{nm} &=& f_{ny} f_{my}.
  \label{P-product}
\end{eqnarray}
The infinite-dimensional matrix $P_{nm}$ is a symmetric one,
and a square of each element of the matrix is integer number.
The first several elements of $P_{nm}$ are calculated as the following:
\newcommand{\msp}{\hspace{2.5mm}}
\begin{eqnarray}
 P_{nm} &=&
 \left[
  \mbox{\scriptsize$
  \begin{array}{r@{\msp}r@{\msp}r@{\msp}r@{\msp}r@{\msp}r@{\msp}r@{\msp}r@{\msp}r@{\msp}rr}
  1& 0& \sqrt{2}& 0& \sqrt{6}& 0& \sqrt{20}& 0& \sqrt{70}& 0&\\
  0& 1& 0& \sqrt{6}& 0& \sqrt{30}& 0& \sqrt{140}& 0& \sqrt{630}&\\
  \sqrt{2}& 0& 3& 0& \sqrt{48}& 0& \sqrt{250}& 0& \sqrt{1260}& 0&\\
  0& \sqrt{6}& 0& 7& 0& \sqrt{320}& 0& \sqrt{1890}& 0& \sqrt{10500}&\\
  \sqrt{6}& 0& \sqrt{48}& 0& 19& 0& \sqrt{2430}& 0& \sqrt{15120}& 0&\\
  0& \sqrt{30}& 0& \sqrt{320}& 0& 51& 0& \sqrt{18522}& 0& \sqrt{121296}&\cdots\\
  \sqrt{20}& 0& \sqrt{250}& 0& \sqrt{2430}& 0& 141& 0& \sqrt{145656}& 0&\\
  0& \sqrt{140}& 0& \sqrt{1890}& 0& \sqrt{18522}& 0& 393& 0& \sqrt{1161288}&\\
  \sqrt{70}& 0& \sqrt{1260}& 0& \sqrt{15120}& 0& \sqrt{145656}& 0& 1107& 0&\\
  0& \sqrt{630}& 0& \sqrt{10500}& 0& \sqrt{121296}& 0& \sqrt{1161288}& 0& 3139&\\
   & & & & & \vdots & & & & & \ddots
  \end{array}
  $}
 \right].\nonumber\\
 \label{Pnm}
\end{eqnarray}

The matrix $P_{nm}$ is invertible.
The proof of the invertibility is given in the next subsection.
By applying the inverse matrix of $P_{nm}$
to both sides of (\ref{P-product}),
we obtain
\begin{eqnarray}
 \delta_{km} &=& (P^{-1})_{kn} f_{ny} f_{my}.
\end{eqnarray}
Thus we find a derivation of the inverse matrix $f^{-1}$ as
\begin{eqnarray}
 (f^{-1})_{ky} &=& (P^{-1})_{kn} f_{ny}.
  \label{f-inverse}
\end{eqnarray}
By substituting (\ref{f-inverse}) into (\ref{g-by-f-inverse}),
we obtain the principal part of the metric tensor
\begin{eqnarray}
 g_{y_1 y_2} &=&
 f_{ny_1} (P^{-1})_{nk} 
  {\cal I}_{kl}
 (P^{-1})_{lm} f_{my_2}.
\end{eqnarray}
Finally, the integral kernel of the inner
product (\ref{G-product}) is obtained as
\begin{eqnarray}
 G(y_1,y_2)
  &=&
  \Lambda(y_1)\Lambda(y_2)\:
  \sum_{n,m}
   \left[ P^{-1} {\cal I} P^{-1} \right]_{nm}
   f_{n}(y_1)
   f_{m}(y_2).
   \label{G-Kernel}
\end{eqnarray}
After giving a proof of the invertibility of the matrix $P_{nm}$
in the next subsection,
we consider the abstract form of the inverse matrix $P^{-1}_{nm}$
in the subsection \ref{Homeomorphism.sec},
and concretely present the inner product
as a reconstruction of the inner product (\ref{G-product})
in the subsection \ref{Reconstruction.sec}.

\subsection{A proof of the invertibility of $P_{nm}$}

We give a proof of the invertibility of $P_{nm}$
by the fact that
any regularized wave function of the negative number sector,
which has been defined in (\ref{f_n}), 
is uniquely expanded by
the wave functions of the positive number sector:
\begin{eqnarray}
 \phi_{+n}(y)
  &\equiv&
  \frac{1}{\pi^{1/4}\sqrt{2^{n}\:n!}} e^{-\frac{1}{2}y^2} H_{+n}(y),
  \label{phi_m_AP}
\end{eqnarray}
where the index takes values $n=0,1,2,\cdots$.

There are recurrence formulae for the function series
$\phi_{+n}$ and $f_{n}$:
\begin{eqnarray}
 \phi_{+n}(y)
  &=&
  \sqrt{\frac{2}{n}} y \phi_{+(n-1)}(y) - \sqrt{\frac{n-1}{n}} \phi_{+(n-2)}(y)
  \label{rec_phi}\\
 f_{n}(y)
  &=&
  \sqrt{\frac{2}{n}} y f_{n-1}(y) + \sqrt{\frac{n-1}{n}} f_{n-2}(y)
  \label{rec_f}
\end{eqnarray}
The norm of $\phi_{+n}$ has been chosen to unity.
We consider the Gram-Schmidt method 
to orthonormalize the function series $f_n(y)$.
We denote the orthonormalized series as $g_n(y)$.
The Gram-Schmidt method is concretely given by
\begin{eqnarray}
 g_n &\equiv& f_n - \sum_{k=0}^{n-1}
  \frac{\langle f_n | g_k \rangle}{\langle g_k | g_k \rangle} g_k.
  \label{GSmethod}
\end{eqnarray}
The first function $g_0(y)$ becomes
\begin{eqnarray}
 g_{0} &=& f_{0} \ =\  \phi_{+0},
\end{eqnarray}
and the second becomes
\begin{eqnarray}
 g_{1} &=& f_{1}
  - \frac{\langle f_1 | g_0 \rangle}{\langle g_0 | g_0 \rangle} g_0.
  \ =\  \phi_{+1}.
\end{eqnarray}
Both $g_{0}$ and $g_{1}$ agree with $\phi_{+0}$ and $\phi_{+1}$
respectively.

Here, we will employ the mathematical induction to prove
$g_n(y) = \phi_{+n}(y)$ for any $n=0,1,2,\cdots$.
If we assume that $g_k(y) = \phi_{+k}(y)$ for $k=0,1,2,\cdots,n-1$,
then the Gram-Schmidt method for $g_n$ becomes the following:
\begin{eqnarray}
 g_n &=& f_n - \sum_{k=0}^{n-1}
  {\langle f_n | \phi_{+k} \rangle} \phi_{+k} \nonumber \\
 &=& f_n-\sqrt{\frac{2}{n}}\> \sum_{k=0}^{n-1}\langle y f_{n-1}|\phi_{+k}\rangle \phi_{+k}-\sqrt{\frac{n-1}{n}}\> \sum_{k=0}^{n-1}\langle f_{n-2}|\phi_{+k}\rangle \phi_{+k}, 
\label{g-sch}
\end{eqnarray}
where we have used the recurrence formulae (\ref{rec_f})
and the orthonormality
$\langle \phi_{+n} | \phi_{+m} \rangle = \delta_{nm}$
of the positive number sector.
The assumption of $g_k(y)$ derives a relation
\begin{eqnarray}
 \phi_{+k}
  &=&
  f_k-\sum_{l=0}^{k-1}\langle f_k | \phi_{+l} \rangle \phi_{+l}
  \label{phi_f_k_GS}
\end{eqnarray}
from the Gram-Schmidt method (\ref{GSmethod}).
By using the recurrence formulae (\ref{rec_phi}) and 
the relation (\ref{phi_f_k_GS}),
the second term of the right-hand side of (\ref{g-sch}) reads 
\begin{eqnarray*}
 \langle y f_{n-1}|\phi_{+k}\rangle 
  &=&
  \langle f_{n-1}|y\phi_{+k} \rangle 
  =
  \langle f_{n-1}|
  \left[
      \sqrt{\frac{k+1}{2}} | \phi_{+(k+1)} \rangle
    + \sqrt{\frac{k}{2}}   | \phi_{+(k-1)} \rangle
  \right]
 \\
 &=&
 \left[
  \langle \phi_{n-1}|
  +\sum_{l=0}^{n-2}\langle f_{n-1}|\phi_{+l}\rangle \langle\phi_{+l}|
 \right]
 \left[
     \sqrt{\frac{k+1}{2}} | \phi_{+(k+1)} \rangle
   + \sqrt{\frac{k}{2}}   | \phi_{+(k-1)} \rangle
 \right]
 \\
 &=&
  \left\{
   \begin{array}{ll}
    \sqrt{\frac{n-1}{2}}\> \langle f_{n-1}|\phi_{+(n-2)}\rangle
     & (k=n-1) \\ 
    \sqrt{\frac{n-1}{2}}
     + \sqrt{\frac{n-2}{2}} \langle f_{n-1}|\phi_{+(n-3)}\rangle
     & (k=n-2) \\
    \sqrt{\frac{k+1}{2}}  \langle f_{n-1}|\phi_{+(k+1)}\rangle 
     + \sqrt{\frac{k}{2}} \langle f_{n-1}|\phi_{+(k-1)}\rangle
     \qquad
     & (k\leq n-3)
   \end{array}\right..
\end{eqnarray*}
By using (\ref{phi_f_k_GS}),
the third term of the right-hand side of (\ref{g-sch}) becomes 
\begin{eqnarray*}
 \langle f_{n-2}|\phi_{+k}\rangle 
  &=&
  \left[
   \langle \phi_{n-2}|
   +
   \sum_{l=0}^{n-3} \langle f_{n-2}|\phi_{+l}\rangle
   \langle \phi_{+l}|
  \right]
  |\phi_{+k}\rangle
  \\
 &=&
  \left\{
   \begin{array}{ll}
    0 & (k=n-1) \\ 
    1 & (k=n-2) \\
    \langle f_{n-2}|\phi_{+k}\rangle\qquad & (k\leq n-3)
   \end{array}\right..
\end{eqnarray*}
By substituting these relations into the relation (\ref{g-sch}),
the relation becomes
\begin{eqnarray}
g_n 
&=& f_n \nonumber\\
&&
 -\sqrt{\frac{2}{n}}
 \left[
   \sqrt{\frac{n-1}{2}}\phi_{+(n-2)}
  +\sum_{k=0}^{n-3}\sqrt{\frac{k+1}{2}} \langle f_{n-1}|\phi_{+(k+1)}\rangle
   \phi_{+k}
  +\sum_{k=1}^{n-1}\sqrt{\frac{k  }{2}} \langle f_{n-1}|\phi_{+(k-1)}\rangle
   \phi_{+k}
 \right]
  \nonumber\\
&&
 -\sqrt{\frac{n-1}{n}}
 \left[
    \phi_{+(n-2)}
  + \sum_{k=0}^{n-3}\langle f_{n-2}|\phi_{+k}\rangle \phi_{+k}
 \right]
  \nonumber\\
&=&
 f_n
 - \sqrt{\frac{n-1}{n}}\> \phi_{+(n-2)}
 - \sqrt{\frac{n-1}{n}}\> f_{n-2} \nonumber\\
&&
 -\sqrt{\frac{2}{n}}
 \left[
  \sum_{k=0}^{n-3}\sqrt{\frac{k+1}{2}}
  \langle f_{n-1}|\phi_{+(k+1)}\rangle
  \phi_{+k}
  +
  \sum_{k=0}^{n-2}\sqrt{\frac{k+1}{2}} \>
  \langle f_{n-1}|\phi_{+k}\rangle \phi_{+(k+1)}
 \right].
 \label{phi_f_k_GS_2}
\end{eqnarray}
By employing the recurrence formulae (\ref{rec_phi}),
the final term of the relation (\ref{phi_f_k_GS_2}) is rewritten into
\begin{eqnarray}
 \sum_{k=0}^{n-2}\sqrt{\frac{k+1}{2}}
  \langle f_{n-1}|\phi_{+k}\rangle \phi_{+(k+1)}
 &=&
 \sum_{k=0}^{n-2}\sqrt{\frac{k+1}{2}}
  \langle f_{n-1}|\phi_{+k}\rangle
  \left[
    \sqrt{\frac{2}{k+1}} y\phi_{+k}
   -\sqrt{\frac{k}{k+1}}\phi_{+(k-1)}
  \right] \nonumber\\
 &=&
  y \sum_{k=0}^{n-2}\langle f_{n-1}|\phi_{+k}\rangle \phi_{+k}
  - \sum_{k=0}^{n-2}\sqrt{\frac{k}{2}}
    \langle f_{n-1}|\phi_{+k}\rangle \phi_{+(k-1)}
    \nonumber\\
 &=&
  y \left(f_{n-1}-\phi_{+(n-1)}\right)
  - \sum_{k=0}^{n-3}\sqrt{\frac{k+1}{2}}
    \langle f_{n-1}|\phi_{+(k+1)}\rangle \phi_{+k}.
    \label{rel_AP}
\end{eqnarray}
By substituting (\ref{rel_AP}) into (\ref{phi_f_k_GS_2}),
we obtain 
\begin{eqnarray}
g_n
&=& f_n-\sqrt{\frac{n-1}{n}}\> \phi_{+(n-2)}-\sqrt{\frac{2}{n}}\> y \left(f_{n-1}-\phi_{+(n-1)}\right)-\sqrt{\frac{n-1}{n}}\> f_{n-2}
\nonumber \\
&=& \sqrt{\frac{2}{n}}\> y \phi_{+(n-1)} - \sqrt{\frac{n-1}{n}} \phi_{+(n-2)}
  \nonumber\\
 &=& \phi_{+n}.
\end{eqnarray}
The here and now, by the mathematical induction,
we have proved
the relation $g_n(y) = \phi_{+n}(y)$ for any $n = 0,1,2,\cdots$.
This means that
the Gram-Schmidt orthonormalized series of the function series $\{f_n(y)\}$
are completely coincident with
the ordinary wave functions $\{\phi_{+n}(y)\}$
of the harmonic oscillator.

Therefore
any function of the series $\{f_n\}$
is uniquely expanded by $\{\phi_{+n}\}$.
Inversely
any function of the series $\{\phi_{+n}\}$
is also uniquely expanded by $\{f_n\}$.
Now we have found the existence of the homeomorphism
among the wave functions $\{\phi_{+n}\}$ of the positive number sector
and
the regularized wave functions $\{f_{n}\}$ of the negative number sector.
The existence of the homeomorphism
implies that the matrix $P_{nm}$ should be invertible.

\subsection{Homeomorphism among negative- and positive-number basis}
\label{Homeomorphism.sec}

We write the homeomorphism
among the wave functions $\{ \phi_{+n} \}$ of the positive number sector
and
the regularized wave functions $\{ f_{n} \}$ of the negative number
sector
as the following matrix form:
\begin{eqnarray}
 \phi_{+n}(y) &=& \sum_{k=0}^{n} A_{nm} f_{m}(y),          \label{PhiF}\\
 f_{n}(y)     &=& \sum_{k=0}^{n} A^{-1}_{nm} \phi_{+m}(y). \label{FPhi}
\end{eqnarray}
The matrix $A_{nm}$ becomes lower triangular,
because the Gram-Schmidt orthonormalization of $\{ f_{n} \}$
derives $\{ \phi_{+n} \}$.
The inverse matrix $A^{-1}_{nm}$ also becomes lower triangular.
By applying the orthonormality
\begin{eqnarray}
 \int dy \phi_{+n}(y) \phi_{+m}(y) &=& \delta_{nm}
 \label{orthonormality_phi}
\end{eqnarray}
to the homeomorphism (\ref{FPhi}),
we find the form of the homeomorphism matrix $A^{-1}_{nm}$ as
\begin{eqnarray}
 A^{-1}_{nm}
  &=& 
 \int_{-\infty}^{+\infty} \: dy {f_n}(y) \phi_{+m}(y).
 \label{AinvDef}
\end{eqnarray}
We concretely drive the matrix elements of $A^{-1}_{nm}$ as
\begin{eqnarray}
 A^{-1}_{nm}
  &=& 
 \left[
  \mbox{\scriptsize$
  \begin{array}{@{}rrrrrrrr}
   1& 0& 0& 0& 0& 0& 0&\\
   0& 1& 0& 0& 0& 0& 0&\\
   \frac{1}{2} + \frac{1}{\sqrt{2}}& 0& 1& 0& 0& 0& 0&\\
   0& \frac{3}{2} + \frac{\sqrt{3}}{\sqrt{2}}& 0& 1& 0& 0& 0&\cdots\\
   \frac{3}{4} + \frac{3\sqrt{2}}{2} + \frac{\sqrt{6}}{4}& 0& 3 + \sqrt{3}& 0& 1& 0& 0&\\
   0& \frac{15}{4} + \frac{5\sqrt{6}}{2} + \frac{\sqrt{30}}{4}& 0& 5 + \sqrt{5}& 0& 1& 0&\\
   \frac{15}{8} + \frac{45\sqrt{2}}{8} + \frac{\sqrt{5}}{4} + \frac{15\sqrt{6}}{8}& 0& \frac{45}{4} + \frac{15\sqrt{3}}{2} + \frac{3\sqrt{10}}{4}& 0& \frac{15}{2} + \frac{\sqrt{15}}{\sqrt{2}}& 0& 1&\\
    &  & \vdots & & & & &  \ddots
  \end{array}
  $}
 \right].\nonumber\\
 \label{AinvMatrix}
\end{eqnarray}
\newcommand{\SF}[2]{{{\frac{\sqrt{#1}}{\sqrt{#2}}}}}%
Each element of the matrix $A_{nm}$
is calculated from (\ref{AinvMatrix})
by finite steps as
\begin{eqnarray}
 A_{nm} &=&
 \left[
  \mbox{\scriptsize$
  \begin{array}{@{}rrrrrrrr}
   1& 0& 0& 0& 0& 0& 0&\\
   0& 1& 0& 0& 0& 0& 0&\\
   -\frac{1}{2} - \frac{1}{\sqrt{2}}& 0& 1& 0& 0& 0& 0&\\
   0& -\frac{3}{2} - \SF{3}{2}& 0& 1& 0& 0& 0&\cdots\\
   \frac{3}{4} + \frac{1}{2}\SF{3}{2} + \frac{1}{2}\sqrt{3}& 0& -3 - \sqrt{3}& 0& 1& 0& 0&\\
   0& \frac{15}{4} + \frac{3}{2}\sqrt{5} + \frac{1}{2}\SF{15}{2}& 0& -5 - \sqrt{5}& 0& 1& 0&\\
   -\frac{15}{8} - \frac{3}{4}\SF{5}{2} - \frac{\sqrt{5}}{4} - \frac{3}{4}\SF{15}{2}& 0& \frac{45}{4} + \frac{3}{2}\SF{5}{2} + 3\SF{15}{2}& 0& -\frac{15}{2} - \SF{15}{2}& 0& 1&\\
    &  & \vdots & & & & &  \ddots
  \end{array}
  $}
 \right].\nonumber\\
 \label{AMatrix}
\end{eqnarray}
All of the diagonal elements of these matrixes (\ref{AinvMatrix}) and
(\ref{AMatrix}) become unity.

By combining 
the definition of the matrix
$P_{nm} := \int dy f_{+n}(y) f_{+m}(y)$ in  (\ref{P-product}),
the homeomorphism (\ref{FPhi}) and
the orthonormality (\ref{orthonormality_phi}),
the matrix $P_{nm}$ is rewritten as
\begin{eqnarray}
  P &=& A^{-1} \cdot (A^{-1})^{\rm T}.
  \label{PAA}
\end{eqnarray}
Therefore the inverse matrix $P^{-1}_{nm}$ can be also rewritten as
\begin{eqnarray}
  P^{-1} &=& A^{\rm T } \cdot A.
  \label{PinvAA}
\end{eqnarray}
We should note that the lower triangular matrixes $A_{nm}$
and $A^{-1}_{nm}$
are calculable by finite steps, however,
the calculation of $P^{-1}_{nm}$ requires infinite steps,
because $P^{-1}_{nm}$ is obtained by the product of the upper
triangular matrix $A^T$ and the lower triangular matrix $A$
in (\ref{PinvAA}).

\subsection{Reconstruction of the Inner Product without $P^{-1}$}
\label{Reconstruction.sec}

Even for the finite calculability of the matrix $A$,
the calculation of $P^{-1}$ in (\ref{PinvAA}) requires infinite steps.
It seems to be better to redefine of the inner product
without using the inverse matrix $P^{-1}_{nm}$.
In this subsection,
we reconstruct the inner product defined in (\ref{G-product})
by using the finite-calculable matrix $A$ instead of the matrix $P^{-1}$.

The integral kernel (metric tensor)
which has been formally derived in (\ref{G-Kernel})
becomes
\begin{eqnarray}
 G(y_1, y_2)
  &=&
 \Lambda(y_1) \Lambda(y_2)
  \sum_{n,m}
   \left[ P^{-1} {\cal I} P^{-1} \right]_{nm}
   f_{n}(y_1)
   f_{m}(y_2)
   \nonumber\\
  &=&
 \Lambda(y_1) \Lambda(y_2)
  \sum_{n,m}
   \left[ A^{\rm T} \cdot A \cdot {\cal I} \cdot A^{\rm T} \cdot A
   \right]_{nm}
   f_{n}(y_1)
   f_{m}(y_2)
   \nonumber\\
  &=&
 \Lambda(y_1) \Lambda(y_2)
  \sum_{n,m}
   \left[ A \cdot {\cal I} \cdot A^{\rm T}
   \right]_{nm}
   (A \cdot f)_{n}(y_1)\;
   (A \cdot f)_{m}(y_2)
   \nonumber\\
  &=&
 \Lambda(y_1) \Lambda(y_2)
  \sum_{n,m}
   \left[ A \cdot {\cal I} \cdot A^{\rm T}
   \right]_{nm}
   \phi_{+n}(y_1)
   \phi_{+m}(y_2),
 \label{G-AP1}
\end{eqnarray}
where we have used the relation (\ref{PhiF}).
We recall the definition of the inner product (\ref{G-product}).
We ignore the positive energy sector,
because the sector is nor relevant in this argument.
The inner product (\ref{G-product}) is rewritten into
\begin{eqnarray}
 \left< f \:|\: g \right>
 &=&
 \int_{-\infty}^{+\infty} dy_1
 \int_{-\infty}^{+\infty} dy_2
 \: G(y_1, y_2)
 \left< f(y_1), \: g(y_2) \right>,\nonumber\\
  &=&
 \int_{-\infty}^{+\infty} dy_1
 \int_{-\infty}^{+\infty} dy_2
 \Lambda(y_1) \Lambda(y_2)
  \sum_{n,m}
   \left[ A \cdot {\cal I} \cdot A^{\rm T}
   \right]_{nm}
   \phi_{+n}(y_1)
   \phi_{+m}(y_2)
   \;
   \left< {f}(y_1), \: {g}(y_2) \right>\nonumber\\
  &=&
   \sum_{n,m}
   \left[ A \cdot {\cal I} \cdot A^{\rm T}
   \right]_{nm}
   \left< 
    \int_{-\infty}^{+\infty} dy_1 \Lambda(y_1) \phi_{+n}(y_1) {f}(y_1),\;
    \int_{-\infty}^{+\infty} dy_2 \Lambda(y_2) \phi_{+m}(y_2) {g}(y_2)
   \right>\nonumber\\
 \label{G-product-AP1}
\end{eqnarray}
Any matrix element of the product $(A \cdot {\cal I} \cdot A^{\rm T})$
in (\ref{G-product-AP1})
is calculated by finite steps,
because the matrix $A$ is lower triangular matrix.
Now we redefine the inner product for the negative number sector
by (\ref{G-product-AP1})
instead of the formal definition (\ref{G-product}) with (\ref{G-Kernel}).

We check the orthonormality of the
non-regularized negative number wave function basis
\begin{eqnarray}
 \phi_{-a}(y)
  &=&
  J^a \otimes
  \frac{1}{\pi^{1/4}\sqrt{2^{a}\:a!}} e^{+\frac{1}{2}y^2} H_{-a}(y),
\end{eqnarray}
where we note that $\phi_{-a}(y) = J^a \otimes f_a(y)/\Lambda(y)$.
The product among the basis becomes
\begin{eqnarray}
  \left< \phi_{-a} \:|\: \phi_{-b} \right>
  &=&
   \sum_{n,m}
   \left[ A \cdot {\cal I} \cdot A^{\rm T}
   \right]_{nm}
    \int_{-\infty}^{+\infty} dy_1 \phi_{+n}(y_1) {f_a}(y_1)
    \int_{-\infty}^{+\infty} dy_2 \phi_{+m}(y_2) {f_b}(y_2)
   \left< J^a, J^b \right>. \nonumber\\
 \label{G-orthonormal-AP1}
\end{eqnarray}
By using the property (\ref{AinvDef}),
%
we find the orthonormality as
\begin{eqnarray}
  \left< \phi_{-a} \:|\: \phi_{-b} \right>
  &=&
   \left[ A^{-1} \cdot A \cdot {\cal I} \cdot A^{\rm T} \cdot (A^{-1})^{\rm T}
   \right]_{ab}
   \left< J^a, J^b \right>. \nonumber\\
  &=&
   {\cal I}_{ab} \left< J^a, J^b \right>
  \ =\ \delta_{ab}.
  \label{G-orthonormal-AP2}
\end{eqnarray}

Finally, we summarize the definition of the inner product
of the double harmonic oscillator system as
\begin{eqnarray}
 \left< f \:|\: g \right>
 &\equiv&
   \int_{-\infty}^{+\infty} dx
   \sum_{n,m}
   \left[ A \cdot {\cal I} \cdot A^{\rm T}
   \right]_{nm} \nonumber\\
 &&
   \qquad\times
   \left< 
    \int_{-\infty}^{+\infty} dy_1 \Lambda(y_1) \phi_{+n}(y_1) {f}(x,y_1),\;
    \int_{-\infty}^{+\infty} dy_2 \Lambda(y_2) \phi_{+m}(y_2) {g}(x,y_2)
   \right>.
   \nonumber\\
 \label{G-product-FIN}
\end{eqnarray}

\section{Mathematical properties of negative number sector}

In this section,
we consider the mathematical properties of negative number sector
by comparing with the positive number ones.
We will see the properties of the function spaces,
and the relations among both sectors.
We will also see
why the non-localness appears in the definition of the inner product.

\subsection{Function spaces of the positive- and negative- number sectors}

\begin{figure}[htbp]
 \begin{center}
  \includegraphics[width=12cm]{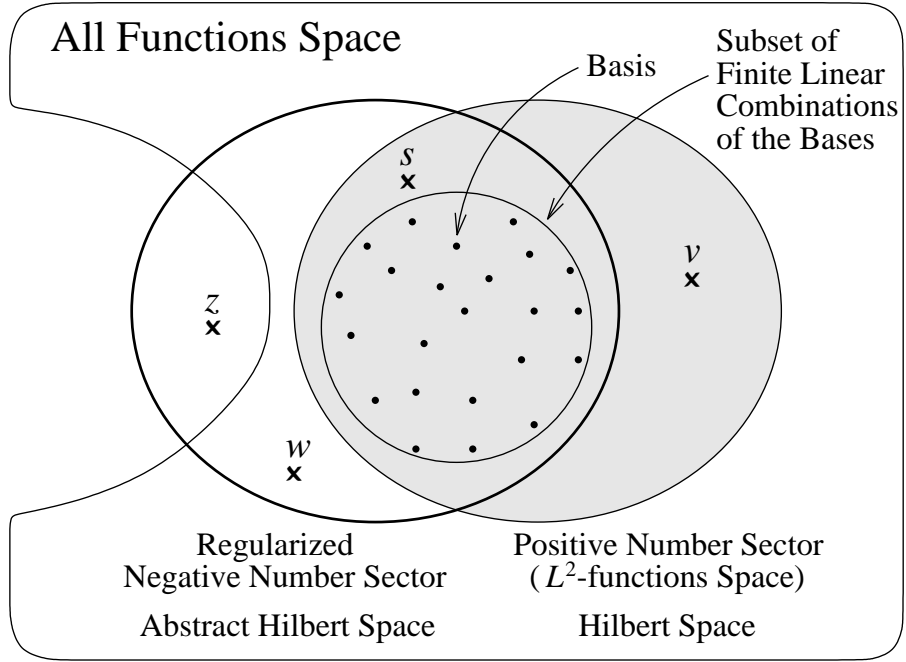}
 \end{center}
 \caption{%
 The space of the functional representations of
 the positive number sector
 and the regularized negative number sector.%
 }
 \label{FuncSpace.eps}
\end{figure}

We summarize the categorization of the functional representations of
the positive number sector $\{\phi_{+n}(y) | n=0,1,2,\cdots \}$
and the regularized negative number sector $\{f_n(y) | n=0,1,2,\cdots \}$
in Fig. \ref{FuncSpace.eps}.
While the positive number sector $\{\phi_{+n}\}$ is originally
defined on $x$-coordinate,
we consider both functions are defined on $y$-coordinate
to compare the properties of these sectors.

Any element of the functional representation of the positive number sector
on $y$-coordinate is given by the linear combination
\begin{eqnarray}
 \phi_{+}(y) &=& \sum_{n=0}^{\infty} c_n \phi_{+n}(y)
\end{eqnarray}
of the bases $\{\phi_{+n}\}$
with the coefficients $\{c_n\}$
which satisfy the condition $\sum_{n=0}^{\infty} |c_n|^2 < \infty$.
It is well-known that
the functional representation of the positive number sector
becomes the square-integrable functions space ($L^2$-functions space),
and also becomes the Hilbert space,
because the $L^2$-functions space includes the limit of any Cauchy sequence.

Any function basis $f_{n}(y)$
of the regularized negative number sector has finite $L^2$-norm,
because
any basis is obtained by the the basis transformation (\ref{FPhi})
from the positive number basis $\{\phi_{+n}(y)\}$,
and the transformation matrix $A^{-1}_{nm}$ in (\ref{FPhi})
is lower triangular.
The metric matrix of $\{f_{n}\}$ defined in (\ref{P-product})
has been concretely calculated in (\ref{Pnm}).
Therefore all bases $\{f_{n}(y)\}$ belong to the $L^2$-functions space.

Any element of the functional representation of
the regularized negative number sector would be given by 
\begin{eqnarray}
 \phi_{-}(y) &=& \sum_{n=0}^{\infty} c_n f_{n}(y)
\end{eqnarray}
of the bases $\{f_{n}\}$ with coefficients $\{c_n\}$
satisfying the condition $\sum_{n=0}^{\infty} |c_n|^2 < \infty$.
Any finite linear combination of the basis $\{f_n(y)\}$
has finite $L^2$-norm,
and thus belongs to the $L^2$-functions space.
The subset of the finite linear combination of the basis $\{f_n(x)\}$
corresponds with that of $\{\phi_n(y)\}$,
because the transformation matrixes $A_{nm}$ and $A^{-1}_{nm}$
exist and are lower triangular.
The subset of the finite linear combination
is dense subset of the $L^2$-function space with respecting
the $L^2$-norm topology.

The (regularizing) inner product (\ref{G-product-AP1})
of the negative number sector
is different from the ordinary inner product
of the positive number sector.
Due to the unbounded of the matrix $A_{nm}$
which is employed in the definition
of the inner product (\ref{G-product-AP1}),
some of the functional representation of the Cauchy series
of the positive number sector
do not belong to the function space of
the regularized negative number sector
(see the element $v$ in Fig. \ref{FuncSpace.eps}).
On the other hand,
some of the functional representation of the Cauchy series
of the regularized negative number sector
do not belong to the $L^2$-function space
(see the element $w$ in Fig. \ref{FuncSpace.eps}).
These properties does not cause any problem,
because we have adopted different inner products in these sectors.

Any Cauchy series in the positive number sector belongs to
the $L^2$-functions space, however, that in the regularized negative number 
sector does not belongs to the $L^2$-functions space always,
because the bases transformation matrix $A^{-1}_{nm}$ is not bounded.
Moreover a part of the Cauchy series do not belong to the functions space,
because there arises divergence in the functional representation
of slowly-converging Cauchy series
(see the element $z$ in Fig \ref{FuncSpace.eps}).
Therefore the functional representation of 
the regularized negative number sector only forms a pre-Hilbert space
rather than the Hilbert space.

The existence of the element $z$
which have no functional representation
seems to be problematic.
Such element of the slowly-converging Cauchy series appears,
only when the wave function or the differentiation of the wave function
is discontinuous,
thus the slowly-converging Cauchy series are not physically important.
The coherent states are physically important,
and are examples of the rapidly-converging Cauchy series.
In the subsection \ref{Coherent.sec},
we will see that coherent states have no divergence in the functional
representation and have good physical interpretation.
In the next subsection,
we construct {\it the mapping operator}
among the positive number sector and negative number one.
By employing the mapping operator,
we can construct a $L^2$-representation of the negative number sector.
Therefore we concludes that
the existence of the slowly-converging Cauchy series is not problem
physically, and is mathematically solved by the mapping operator.

\subsection{Mapping operators among the positive and negative sector}

\newcommand{\LL}{\langle\hspace{-0.2em}\langle}
\newcommand{\RR}{\rangle\hspace{-0.2em}\rangle}

We consider the relation among the positive and negative number sector
as abstract linear space.
In this subsection, the product $\langle f | g \rangle$
is the ordinary inner product ($L^2$-product) for $f$ and $g$,
and $\LL f | g \RR$ is the inner product
which has been defined in (\ref{G-product}) or in (\ref{G-product-FIN}).
Let $|\phi_{\pm n}\rangle$ be a number state of the positive and
negative number sectors respectively.
Let $\langle\phi_{+n}|$ be a linear functional such that
\begin{eqnarray}
 \langle\phi_{+n}|: &\;& |\phi_{+m}\rangle
 \mapsto \langle\phi_{+n}|\phi_{+m}\rangle = \delta_{nm},
\end{eqnarray}
namely the dual element of $|\phi_{+n}\rangle$,
which is respecting the inner product $\langle|\rangle$
of the positive number sector.
We also define the linear functional $\LL\phi_{-n}|$ such that
\begin{eqnarray}
 \;
 \LL\phi_{-n}| : &\;& |\phi_{-m}\rangle
 \mapsto
 \LL\phi_{-n}|\phi_{-m}\RR = \delta_{nm},
\end{eqnarray}
namely the dual element of $|\phi_{-n}\rangle$,
which is respecting the inner product $\LL|\RR$
of the negative number sector.

Here we define a mapping operator:
\begin{eqnarray}
 {\cal M}
  &\equiv&
  \sum_{n=0}^{\infty} |\phi_{+n}\rangle \otimes \LL \phi_{-n}|,
  \label{mapping1}
\end{eqnarray}
which transforms the negative number particle
to the corresponding positive number one.
The inverse of (\ref{mapping1}) is easily derived as
\begin{eqnarray}
 {\cal M}^{-1}
  &\equiv&
  \sum_{n=0}^{\infty} |\phi_{-n}\rangle \otimes \langle \phi_{+n}|,
  \label{mapping2}
\end{eqnarray}
which transforms the positive number particle
to the corresponding negative number one.
These mapping operators works like the following:
\begin{eqnarray}
 {\cal M} |\phi_{-n}\rangle \ =\ |\phi_{+n}\rangle, &\qquad&
 {\cal M}^{-1} |\phi_{+n}\rangle \ =\ |\phi_{-n}\rangle.
\end{eqnarray}

We denote the ordinary positional state as $| x \rangle$,
which satisfies
\begin{eqnarray}
 \langle x | y \rangle = \delta(x-y). \label{xydelta}
\end{eqnarray}
The functional representations of the states $|\phi_{\pm n} \rangle$,
namely, the wave functions are given by
\begin{eqnarray}
 \phi_{\pm n}(x) &=& \langle x | \phi_{\pm n} \rangle.
\end{eqnarray}
We introduce a corrective positional state
for the negative number sector by
\begin{eqnarray}
 | y^c \rangle &\equiv& {\cal M}^{-1} | y \rangle 
 \ =\ \sum_{n=0}^{\infty} | \phi_{-n} \rangle
    \;\langle \phi_{+n} | y \rangle
 \ =\ \sum_{n=0}^{\infty} | \phi_{-n} \rangle \; \phi^*_{+n}(y).
\end{eqnarray}
The wave function of the corrective positional state $| y^c \rangle $
becomes
\begin{eqnarray}
 \langle x | y^c \rangle
  &=&  \sum_{n=0}^{\infty}
      \langle x         | \phi_{-n} \rangle
    \;\langle \phi_{+n} | y         \rangle
 \ =\ \sum_{n=0}^{\infty} \phi_{-n}(x) \; \phi^*_{+n}(y).
 \label{xynonlocal}
\end{eqnarray}
By comparing with (\ref{xydelta}),
the state $| y^c \rangle $ has non-local wave function.
The duals of $| y^c \rangle $ are calculated as
\begin{eqnarray}
 \langle y^c | &=&
  \sum_{n=0}^{\infty}
  \langle y | \phi_{+n} \rangle \; \langle \phi_{-n} |,\\
 \LL y^c | &=&
  \sum_{n=0}^{\infty}
  \langle y | \phi_{+n} \rangle \; \LL \phi_{-n} |
  \ =\ \langle y| {\cal M},
\end{eqnarray}
and they satisfy $\langle x^c | y^c \rangle = \delta(x-y)$ and
$\LL x^c | y^c \RR = \delta(x-y)$.

Here we find the relation:
\begin{eqnarray}
 \LL y^c | \phi_{-n} \RR
  &=&  \langle y | {\cal M} | \phi_{-n}\rangle
  \ =\ \langle y | \phi_{+n}\rangle
  \ =\ \phi_{+n}(y).
\end{eqnarray}
This relation implies that
the wave function $\LL y^c | \phi_{-n} \RR$ of the negative number sector
with respecting the inner product $\LL|\RR$
and the corrective positional state $|y^c\rangle$
completely agrees with the wave function $\phi_{+n}(y)$
of the positive number sector.
Therefore
we can completely construct the Hilbert space of the negative number
sector by the wave function with respecting
the the inner product $\LL|\RR$ and the corrective positional state
$|y^c\rangle$.
Finally, the mapping operators ${\cal M}$ and ${\cal M}^{-1}$
give us the homeomorphism of the Hilbert space among positive and
negative number sector.

The relation among the ordinary positional state $| x \rangle$
and the corrective positional state $| y^c \rangle$ for the
negative number sector is non-local in (\ref{xynonlocal}).
This is the reason why the non-local inner product (\ref{G-product})
arises in the theory.

\subsection{Consistency of numerical calculations with finite matrix size}

We have temporary used the integral kernel
$G(y_1,y_2)$ derived in (\ref{G-Kernel})
to construct the final form (\ref{G-product-FIN}) of the inner product
for the double harmonic oscillator system,
and the integral kernel (\ref{G-Kernel}) is not required
in the final definition (\ref{G-product-FIN}).
The integral kernel $G(y_1,y_2)$ 
plays the metric tensor of the wave-functional space,
and its non-diagonal part indicates the non-localness
of the inner product (\ref{G-product}).
It would be interesting to show the concrete function form
of the integral kernel.
We have obtained the abstract form of the inverse matrix $P^{-1}_{nm}$
in (\ref{PinvAA})
to derive the integral kernel (\ref{G-Kernel}),
however,
the calculation of each element of $P^{-1}_{nm}$ requires infinite steps.

In this subsection, we show that
the finite cut-off of the infinite matrix $P_{nm}$ 
is consistent for any given size $N$,
and present the concrete function form of the integral kernel
$G(y_1,y_2)$ for finite $N$.
This consistency of the finite calculations is quite desirable,
when we consider numerical calculations, approximations and so on.

We define the finite submatrix $\bar{P}_{nm}$ of $P_{nm}$
with $n,m=0,1,\cdots, N$,
where $N$ is the arbitrary finite integer.
We also define the finite submatrix $\bar{A}_{nm}$
of $A_{nm}$ with $n,m=0,1,\cdots, N$.
For any given $N$,
the determinant of the submatrix $\bar{A}_{nm}$ becomes
\begin{eqnarray}
 \det\bar{A} &=& 1.
 \label{detA}
\end{eqnarray}
because the matrix $\bar{A}_{nm}$ is triangular,
and each of its diagonal elements is unity.
Thus the submatrix $\bar{A}_{nm}$ for any given $N$ is always invertible.
The submatrixes can be written as
\begin{eqnarray}
  \bar{P} &=& \bar{A}^{-1} \cdot (\bar{A}^{-1})^{\rm T},
  \qquad
  \bar{P}^{-1} \ =\  \bar{A}^{\rm T } \cdot \bar{A},
  \label{PAAbar}
\end{eqnarray}
and their determinant becomes also unity:
\begin{eqnarray}
 \det\bar{P} \ =\  \det\bar{P}^{-1} &=& 1,
  \label{detbarP}
\end{eqnarray}
which are independent of $N$.

The invertibility of the finite matrix $\bar{P}_{nm}$
and the property in (\ref{detbarP})
are quite desirable,
because these properties assure the consistency of
the numerical calculation for any finite $N$.

\begin{figure}[ht]
 \begin{center}
  \includegraphics[width=80mm]{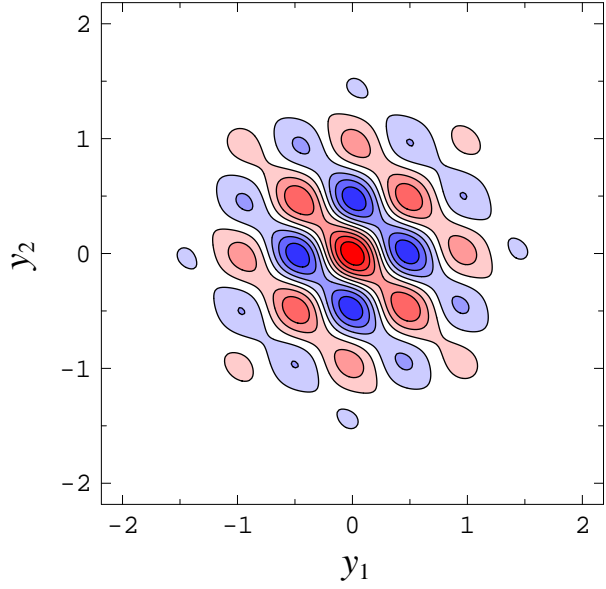}
 \end{center}
 \caption{%
 Contour graph of the integral kernel $G(y_1,y_2)$ for $N=20$.
 This is color graphics.
 The red and the blue indicate positive and negative values
 respectively.
 }
 \label{G-func}
\end{figure}

We have calculated the numerical form of the integral kernel
$G(y_1,y_2)$ for $N=20$ as for example,
and it is drawn in Fig.~\ref{G-func}.
As results of the numerical calculations for several values of $N$,
the shape of $G(y_1,y_2)$ becomes minuter when $N$ increases.
When we consider the limit $N\rightarrow\infty$,
the integral kernel $G(y_1,y_2)$ seems to become hyper-function
rather than ordinary function.

\section{The physical meanings of boson vacuum}

In this section,
we consider the physical meanings of the boson vacuum
in the negative number sector.

\subsection{Fermionic harmonic oscillator}

We review the first quantization of the fermionic harmonic oscillator
to present the basics of the Dirac sea for fermions.
We can separate the positive number sector and negative number one
in the argument,
because the ordinary definition of the inner product works properly
and no regularization of the product is needed.

The one-dimensional fermionic harmonic oscillator
is described by the Grassmann odd operators
which satisfy the anti-commutator
\begin{align}
	\{b,b^{\dagger}\}=1. \label{3.38}
\end{align}
These operators are represented
in terms of a real Grassmann variable $\theta$ as 
\begin{align}
	b=\theta, \qquad b^{\dagger}=\frac{d}{d\theta}.
	\label{3.39}
\end{align}
%
%
The Hamiltonian is given by
\begin{align}
	H=\frac{1}{2}[b^{\dagger},b]=b^{\dagger}b-\frac{1}{2}=N-\frac{1}{2}, 
	\label{3.40}
\end{align}
where $N \equiv b^{\dagger}b$ is the number operator.
The Schr\"{o}dinger equation 
\begin{align}
	\left(b^{\dagger}b-\frac{1}{2}\right)|\tilde{n}_+\rangle 
	=E_{n_+} |\tilde{n}_+\rangle
	\label{3.41}
\end{align}
gives the following solutions:
\begin{eqnarray}
	&&
	E_{n_+} = n-\frac{1}{2},\qquad n=0,1,\\
	&&
	\rho_{0_+}(\theta) = \theta \simeq |\tilde{0}_+\rangle, \qquad
	\rho_{1_+}(\theta) = 1 \simeq |\tilde{1}_+\rangle,
\end{eqnarray}
where the normalization condition 
\begin{align}
	\int d\theta \>
	\rho_{n_+}^{\ast}(\theta)\left(b+b^{\dagger}\right)
	\rho_{n_+'}(\theta)=\delta_{nn'}
	\label{3.46}
\end{align}
has been used.
The vacuum state $|0_+\rangle$ vanishes
by the annihilation operator $b$ as
\begin{eqnarray}
	b|\tilde{0}_+\rangle &=& 0.
	\label{3.45}
\end{eqnarray}

According to the argument in Section 2,
the negative number sector of fermion is derived by
the exchanges
$b \rightarrow d^\dagger$ and
$b^\dagger \rightarrow d$.
The Hamiltonian for negative number sector becomes
\begin{align}
	H=\frac{1}{2}[d,d^{\dagger}]=dd^{\dagger}-\frac{1}{2},
	\label{3.47}
\end{align}
and consequently the Schr\"{o}dinger equation becomes
\begin{align}
	\left(dd^{\dagger}-\frac{1}{2}\right)
	|-\tilde{m}_-\rangle 
	= E_{-m_-}|-\tilde{m}_-\rangle.
	\label{3.48}
\end{align}
By introducing a new real Grassmann variable $\varphi$
and representation of the operators:
$d=\varphi$ and $d^{\dagger}=\frac{d}{d\varphi}$,
the solutions of (\ref{3.48}) are given by 
\begin{align}
	& E_{-m_-} = m - \frac{1}{2},\qquad m=0,1 \\
	&
	\chi_{0_-}(\varphi)   = 1       \simeq |\tilde{0}_-\rangle, \qquad
	\chi_{-1_-}(\varphi)  = \varphi \simeq |-\tilde{1}_-\rangle,
	\label{negative_fermion_wavefunc}
\end{align}
where we have used the same normalization condition as (\ref{3.46}).
The vacuum $|\tilde{0}_-\rangle$ in the negative number sector
is annihilated by the creation operator $d^{\dagger}$ as
\begin{eqnarray}
	d^{\dagger}|\tilde{0}_-\rangle &=&0.
	\label{3.52}
\end{eqnarray}

By considering an empty state $\left|\widetilde{\emp}_-\right>$
of the negative number sector,
the vacuum $\left|\tilde{0}_-\right>$
is identified as the Dirac sea.
The empty state satisfies $d \left|\widetilde{\emp}_-\right> = 0$,
so that we find $\left|\widetilde{\emp}\right> \simeq \varphi$.
The vacuum is derived from the empty state as
\begin{align}
	& |\tilde{0}_-\rangle = d^{\dagger}|\widetilde{\emp}_-\rangle,
	\label{3.53}
\end{align}
due to (\ref{negative_fermion_wavefunc}).
We can thus consider that
the vacuum $|\tilde{0}_-\rangle$ consists of a single quantum
on the empty state $|\widetilde{\emp}_+\rangle$.
Therefore,
the vacuum $|\tilde{0}_-\rangle$ describes the filled state,
and we can construct the Dirac sea in the second quantization of fermions
by using this vacuum $|\tilde{0}_-\rangle$.

\subsection{The meaning of boson vacuum in the negative number sector}

By applying the arguments of fermions in the previous subsection
to bosons,
we clarify the physical meaning of the boson vacuum
$| {0}_+, {0}_-\rangle
\simeq e^{-\frac{1}{2}x^2 +\frac{1}{2}y^2 }$
as the boson sea.
We only consider the vacuum of the negative number sector,
\begin{eqnarray}
 | {0}_-\rangle &\simeq& e^{+\frac{1}{2}y^2 },
\end{eqnarray}
because we do not need to refer the positive number sector
and the inner product in this argument.

According to the previous subsection,
we define an empty state $\left|{\emp}_-\right>$
of the negative number sector.
It is natural to define the empty state as
\begin{eqnarray}
	|{\emp}_-\rangle &\simeq& e^{-\frac{1}{2}y^2},
	\label{3.55}
\end{eqnarray}
because the empty state may have similar form
to the ordinary vacuum $| {0}_+\rangle \simeq e^{-\frac{1}{2}x^2}$
of the positive number sector.
The state $|{\emp}_-\rangle$
is just empty of the negative number quantum,
because this empty state is annihilated by the annihilation operator as
$a_- |{\emp}_-\rangle = 0$.

By using the simple equation
\begin{eqnarray*}
 e^{+\frac{1}{2}y^2} &=& e^{y^2}\cdot e^{-\frac{1}{2}y^2},
\end{eqnarray*}
we find the relation
\begin{align}
	|{0}_-\rangle
	= e^{\frac{1}{2}(a_-+a_-^{\dagger})^2}|{\emp}_-\rangle,
	\label{3.56}
\end{align}
which is nothing but
a bosonic version of the fermionic relation (\ref{3.53}).
This relation indicates that
the vacuum $|0_-\rangle$ is a kind of the coherent state
constructed on the empty state $|\emp_-\rangle$.
All of the even number states of the vacuum
have a non-zero coefficient on the empty state,
because of the operation of the exponent $(a_{-} + a_{-}^{\dagger})^2$
on the empty state in (\ref{3.56}).
The relation (\ref{3.56}) in fact suggests
that the boson vacuum $\left| 0_- \right>$
includes infinite number of quanta in the negative number sector,
because
the creation operator $a_-^\dagger$
operates on the empty state infinite times.
The fermion vacuum $|\tilde{0}_-\rangle$ consists of a single quantum
on the fermionic empty state $|\widetilde{\emp}_-\rangle$,
while
the boson vacuum $|0_-\rangle$ consists of infinite number of quanta
on the bosonic empty state $|{\emp}_-\rangle$.
It is natural that
the boson vacuum includes infinite number of quanta
in the negative number sector,
because there is no exclusion principle for bosons.
Therefore, the boson vacuum $|0_-\rangle$
can be regarded as a kind of filled-state.
We refer to the boson vacuum $|0_-\rangle$ as {\it the boson sea}.
In the second quantization theory,
the boson vacuum $|0_-\rangle$ describes the boson sea
like
the fermion vacuum $|\tilde{0}_-\rangle$ describes the Dirac sea.

A quantum which is created by the annihilation operator $a_-$
on the boson vacuum $\left| 0_- \right>$
is just identified as {\it a bosonic hole} in {the boson sea}.
The number operator of the hole is given by
\begin{eqnarray}
 N_{\rm h^-} &=& a_{-} a_{-}^{\dagger},
\end{eqnarray}
because the hole is created by $a_{-}$.
Its expectation value for the boson sea becomes
\begin{eqnarray}
  {\langle 0_- | N_{\rm h^-} | 0_-\rangle}
  &=& 0,
 \label{B4.25}
\end{eqnarray}
since $a_{-}^{\dagger} |0_-\rangle=0$.
On the other hands,
the absolute value of the particle-number
in the negative number sector is counted by
\begin{eqnarray}
 N_{\rm p^-} &=& a_{-}^{\dagger} a_{-}.
\end{eqnarray}
Its expectation value for empty state 
becomes zero as $\langle \emp_- | N_{\rm p^-} | \emp_-\rangle = 0$.

The inner product,
which is defined by
(\ref{G-product}) or (\ref{G-product-FIN}) 
in the previous section,
gives us
\begin{eqnarray}
	{\langle 0_+,0_-| N_{\rm p^-} |0_+,0_-\rangle}
	&=&
	1.
	\label{NpExpValue}
\end{eqnarray}
This result is consistent with the algebraic relation:
\begin{eqnarray*}
 \langle 0_+,0_-| N_{\rm p^-} |0_+,0_-\rangle
  &=&
  \langle 0_+,0_-| a_{-}^{\dagger} a_{-} |0_+,0_-\rangle
  \;=\;
  \langle 0_+,-1_- | 0_+,-1_-\rangle = 1.
\end{eqnarray*}
The result (\ref{NpExpValue}) from the regularization
suggests that the boson sea is filled by one quantum on the average.
This property is preferable for the supersymmetry,
because the Dirac sea is also filled by a single fermionic quantum.

\subsection{Coherent states of the negative number sector}
\label{Coherent.sec}

One of the most important states of the harmonic oscillator
is the coherent state,
which is one of the rapidly-converging Cauchy series.
In this subsection,
we consider the coherent states of the negative number sector,
and find that the wave function of the coherent states is well defined.
The resultant wave function
seems to be describing the behavior of coherent motion of a hole.

\begin{figure}[ht]
 \begin{center}
  \includegraphics{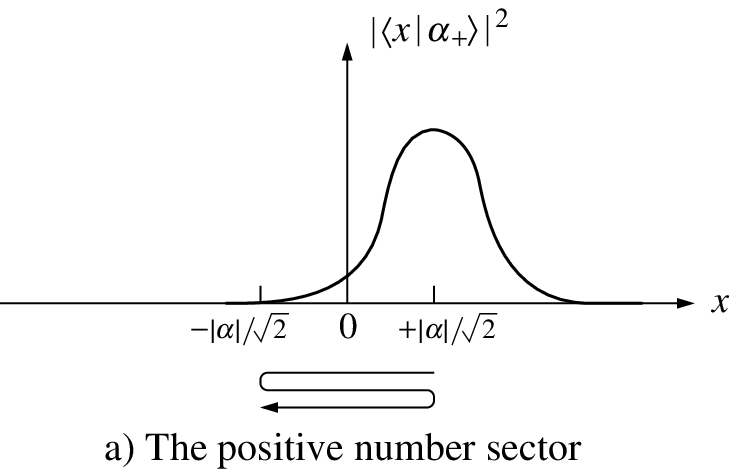}
  \hspace{2em}
  \includegraphics{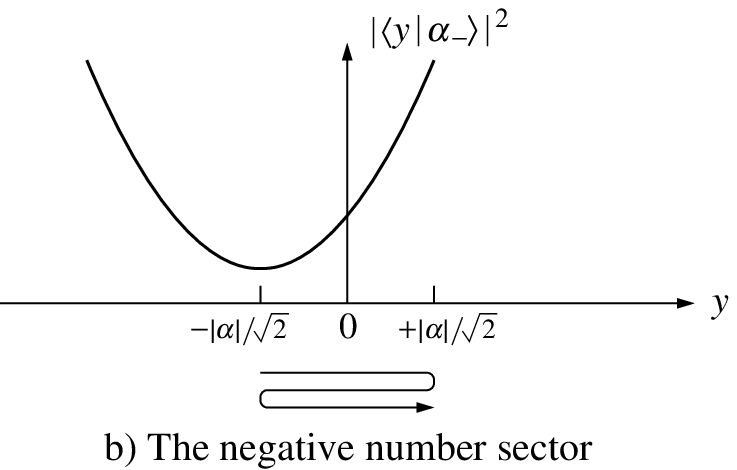}
 \end{center}
 \caption{Distributions of the coherent states.}
 \label{coherent}
\end{figure}

The coherent state $|\alpha_{+}\rangle$
of the ordinary harmonic oscillator is defined by
the eigenstate of the annihilation operator:
\begin{eqnarray}
 a |\alpha_{+}\rangle &=& \alpha |\alpha_{+}\rangle.
\end{eqnarray}
The coherent state is labeled by a complex parameter $\alpha$,
which is the eigenvalue of the annihilation operator.
The time evolution of the probability density
$| \langle x |\alpha_{+}\rangle  |^2$
describes that
the minimum-uncertainty Gaussian wave-packet has reciprocating motion
around the center $x=0$ with the frequency $\omega$.
The reciprocating motion of the wave-packet
is schematically shown in Fig.~\ref{coherent}-a.
The amplitude of the reciprocating motion of the
maximum position becomes $|\alpha|/\sqrt{2}$.

The coherent state $|\alpha_{-}\rangle$
of the negative number sector is defined by
the eigenstate of the creation operator:
\begin{eqnarray}
 a^{\dagger} |\alpha_{-}\rangle &=& \alpha |\alpha_{-}\rangle.
\end{eqnarray}
The coherent state of the negative number sector is also
labeled by a complex parameter $\alpha$,
which is the eigenvalue of the creation operator.
The time evolution of $| \langle y |\alpha_{-}\rangle |^2$
describes that
the inverted Gaussian function has reciprocating motion
around the center $y=0$ with the frequency $\omega$
(see Fig.~\ref{coherent}-b).
The position of the minimum of the inverted Gaussian function
is reciprocating, and the amplitude of the reciprocating motion of the
minimum position is $|\alpha|/\sqrt{2}$.
The amplitude $|\alpha|/\sqrt{2}$ corresponds with
that of the positive number sector.
The phase of the maximum of $| \langle x |\alpha_{+}\rangle |^2$
is different of $\pi$
from that of the minimum in $| \langle y |\alpha_{-}\rangle |^2$.
The phase difference of $\pi$ may come from the fact that
any sign of the quantum numbers in the negative number sector
is inverted from that in the positive number one.

The resultant behavior of the coherent state
of the negative number sector seems to be describing
the coherent motion of a hole with the minimum uncertainly.
This is quite desirable for physical picture
of the negative number sector.

\section{Boson sea}

In the present section, 
we apply the method of the double harmonic oscillator
to the second quantization of complex scalar fields,
for which we investigate the structure of the boson sea.

The bosonic part of the system
corresponds to the double harmonic oscillators of the infinite number.
The double harmonic oscillators are labeled
by the flavor index $i$ and the momentum $\vec{k}$.
We introduce parameter-functions $X_i(\vec{k})$ and $Y_i(\vec{k})$
which correspond to the parameters $x$ and $y$
of the double harmonic oscillator respectively.
We define the creation and annihilation operators
for complex scalar bosons $A_i$ as
\begin{eqnarray}
 a_{i+}(\vec{k})
  &=&
  I \otimes \frac{1}{\sqrt{2}}
  \left( X_{i}(\vec{k})+\frac{\delta}{\delta X_{i}(\vec{k})} \right), 
  \nonumber \\
 a_{i+}^{\dagger}(\vec{k})
  &=&
  I \otimes \frac{1}{\sqrt{2}}
  \left( X_{i}(\vec{k})-\frac{\delta}{\delta X_{i}(\vec{k})} \right),
  \\
 a_{i-}(\vec{k})
  &=& J \otimes \frac{1}{\sqrt{2}}
  \left( Y_{i}(\vec{k})+\frac{\delta}{\delta Y_{i}(\vec{k})} \right), 
  \nonumber \\
 a_{i-}^{\dagger}(\vec{k})
  &=& J \otimes \frac{1}{\sqrt{2}}
  \left( Y_{i}(\vec{k})-\frac{\delta}{\delta Y_{i}(\vec{k})} \right).
  \label{4.7}
\end{eqnarray}
The operators $a_{i+}(\vec{k})$ and $a_{i+}^{\dagger}(\vec{k})$
satisfy the bosonic algebra (\ref{B+com}) of the positive energy,
and the operators $a_{i-}(\vec{k})$ and $a_{i-}^{\dagger}(\vec{k})$
satisfy the bosonic algebra (\ref{B-com}) of the negative energy.
The bosonic part of the Hamiltonian becomes
\begin{eqnarray}
 H
  &=& \sum_i \int \frac{d^3\vec{k}}{(2\pi )^3}
  \left\{
   \left(
    -\frac{1}{2} \frac{\delta^2}{\delta X_{i}^2(\vec{k})}
    +\frac{1}{2} X_{i}^2(\vec{k})
   \right)
   -
   \left(
    -\frac{1}{2} \frac{\delta^2}{\delta Y_{i}^2(\vec{k})}
    +\frac{1}{2} Y_{i}^2(\vec{k})
   \right)
  \right\},\ \ 
\end{eqnarray}
and the Schr\"{o}dinger equation is
\begin{eqnarray}
	H \Phi[X, Y]=E \Phi [X, Y],
\end{eqnarray}
where $\Phi [X,Y]$ denotes a wave functional of the system.
We are now able to write explicitly a wave functional
for the bosonic vacuum:
\begin{eqnarray}
 \Phi_{\rm vac} [X,Y] &=&
  \exp
  \left[
   - \frac{1}{2} \sum_i \int \frac{d^3\vec{k}}{(2\pi )^3}
   \left(
    X_{i}^2 (\vec{k}) - Y_{i}^2 (\vec{k})
   \right)
  \right],
\end{eqnarray}
which just corresponds to
the bosonic part $\left||0_+\right> \otimes \left||0_-\right>$
of the vacuum (\ref{2.23}).

\subsection{Property of the vacuum}

The creation operator $a_{i+}^{\dagger} (\vec{k})$
creates the ordinary particle with the momentum
$k_\mu = (k_0,\: \vec{k} )$,
where $k_0 = \sqrt{|\vec{k}|^2+m^2}$ is energy of the particle.
The annihilation operator $a_{i-}(\vec{k})$
creates a hole of the momentum $k_\mu$ in the boson sea.
The hole is identified as an anti-particle in the theory .
The number operator of the hole is
\begin{eqnarray}
 N_{i \rm h^-}(\vec{k})
 &=&
 a_{i-}(\vec{k}) a_{i-}^{\dagger}(\vec{k}),
 \label{4.24}
\end{eqnarray}
and its expectation value for the sea becomes
\begin{align}
 {\langle 0 || N_{i \rm h^-}(\vec{k}) || 0 \rangle}
 =0,
 \label{4.25}
\end{align}
since $a_{i-}^{\dagger}(\vec{k})||0_-\rangle=0$.

In contrasted with (\ref{4.24}),
the number operator of the negative energy particles is given by
\begin{align}
 N_{i \rm p^-}(\vec{k}) = a_{i-}^{\dagger}(\vec{k}) a_{i-}(\vec{k}),
 \label{4.26}
\end{align}
since the negative energy particle is created
by the creation operator $a_{i-}^{\dagger}(\vec{k})$.
The vacuum expectation value for given $i$ and $\vec{k}$ becomes
\begin{eqnarray}
 {\langle 0||N_{i\rm p^-}(\vec{k})||0\rangle}
  &=&
   \left\{
    \begin{array}{ll}
      \displaystyle
      \infty
      \quad
      & \mbox{(without any regularization)}\\[0.8em]
      1
      & \mbox{(with the regularization)}
    \end{array}
   \right..
\end{eqnarray}
Without any regularization,
e.g., the naive product in (\ref{naive-product}),
it seems that the sea contains the infinite number 
of the negative energy particles at each negative energy modes. 
The proper regularization results
that the sea is filled up by {\it one} particle on average at each modes.

\begin{figure}[htbp]
 \begin{center}
  \includegraphics{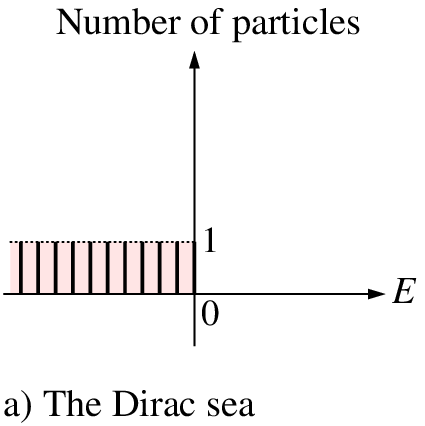}
  \hspace{5em}
  \includegraphics{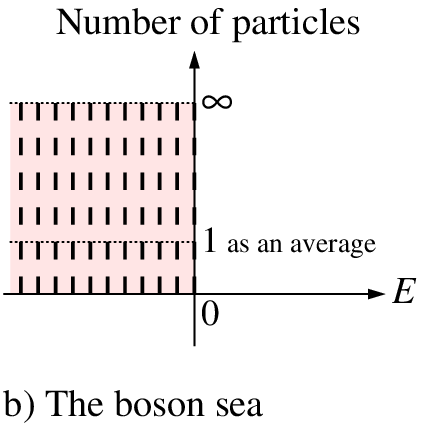}
 \end{center}
 \caption[vacua]{Dirac sea and Boson sea}
 \label{vacua}
\end{figure}

In summary,
the situations of the Dirac sea and the boson sea
may be drawn in Fig.~\ref{vacua}.
In the Dirac sea,
each negative energy level
is occupied by one negative number particles.
This situation is stable due to the exclusion principle.
In the boson sea without regularization,
each negative energy level
seems to be filled by infinitely many negative number particles.
In the boson sea with the proper regularization,
the levels are filled by one negative number particles on the average.

\begin{figure}[htbp]
 \begin{center}
  \includegraphics{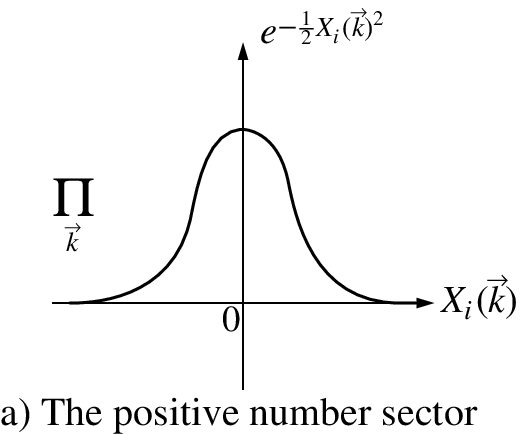}
  \hspace{3em}
  \includegraphics{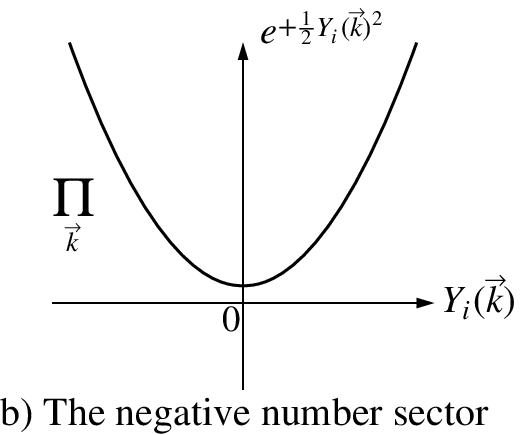}
 \end{center}
 \caption{Wave functionals of the vacuum}
 \label{gaussField}
\end{figure}

In terms of the wave function of second quantization,
we can regard the vacuum $||0_-\rangle$ as the boson sea
in a different way.
In the first quantization language,
the square of the wave function $X_{i}(\vec{k})$ or $Y_{i}(\vec{k})$ 
represents a probability that the particle has the momentum $\vec{k}$.
The probability means the number of the particles
in the second quantization.
The vacuum of the positive energy sector
has a strong peak at $X_{i}=0$,
because the wave functional $\Phi$ is multiplication of
Gaussian functions for any $\vec{k}$ (see in Fig.~\ref{gaussField}-a).
Therefore, 
the configuration that any mode has almost no particles
is dominant in the vacuum.
On the other hand,
the vacuum $||0_-\rangle$ of the negative energy sector
has a strong ``peak'' at $Y_{i}=\pm \infty$ (see in Fig.~\ref{gaussField}-b),
then
the dominant situation is that 
the particle numbers of any mode are infinite.

\subsection{Property of the exited states}

We study the exited states, namely, the particles and the holes
by considering the energy and momentum of the states.
In this subsection
we omit the subscript of the flavor index $i$ for simplicity.
The momentum operator is derived by Noether's theorem: 
\begin{align}
	P^{\mu}
	&=
	\int d^3\vec{x} \: \left\{
	\frac{\partial \mathcal{L}}
	{\partial (\partial_{\mu}A(x))}\partial_0A(x)
	+\frac{\partial \mathcal{L}}{\partial (\partial_{\mu}A^{\dagger}(x))}
	\partial_0A^{\dagger}(x)-\eta^{\mu 0}\mathcal{L}
	\right\} 
	\nonumber \\
	&=
	\int d^3\vec{k}\> \frac{1}{2}k^{\mu}
	\left\{
	 a_{+}
	 a_{+}^{\dagger}(\vec{k})
	+a_{+}^{\dagger}
	 a_{+}(\vec{k})
	+a_{-}
	 a_{-}^{\dagger}(\vec{k})
	+a_{-}^{\dagger}
	 a_{-}(\vec{k})\right\}.
	\label{4.13}
\end{align}
By using the commutation relations (\ref{B+com})  and (\ref{B-com}),
the momentum operator becomes
\begin{align}
	P^{\mu}
	=
	\int d^3\vec{k} \: k^{\mu}
	\left\{
	 a_{+}^{\dagger}
	 a_{+}(\vec{k})
	+a_{-}
	 a_{-}^{\dagger}(\vec{k})
	\;+\;
	\int \frac{d^3\vec{x}}{(2\pi)^3}\> 1\right\}.
	\label{4.14}
\end{align}

We calculate energy and momentum of the vacuum
$||0\rangle_{\text{boson}}=||0_+\rangle \otimes ||0_-\rangle$
as eigenvalues:
\begin{eqnarray}
	P^{\mu}||0\rangle_{\text{boson}}
	&=&
	 \left\{
	  \int \frac{d^3\vec{k}\:d^3\vec{x}}{(2\pi)^3} \: k^{\mu}
	 \right\}
	 ||0\rangle_{\text{boson}},
	 \label{4.15}
\end{eqnarray}
which indicates that the vacuum energy $P^0$ is divergent.
When we have a supersymmetry,
the bosonic vacuum energy in (\ref{4.15})
is exactly canceled by the fermionic one.
The energy-momenta of the first excited states are calculated similarly: 
\begin{eqnarray}
	P^{\mu} a_{+}^{\dagger}(\vec{p}) ||0\rangle_{\text{boson}} 
	&=&
	\left\{
	 p^{\mu}+\int \frac{d^3\vec{k}\:d^3\vec{x}}{(2\pi)^3}\> k^{\mu}
	\right\}
	a_{+}^{\dagger}(\vec{p})||0\rangle_{\text{boson}},\\
	P^{\mu}a_{-}(\vec{p}) ||0\rangle_{\text{boson}}
	&=&
	\left\{
	 p^{\mu}+\int \frac{d^3\vec{k}\:d^3\vec{x}}{(2\pi)^3}\> k^{\mu}
	\right\}
	a_{-}(\vec{p}) ||0\rangle_{\text{boson}}.
	\label{4.17}
\end{eqnarray}
These results tell us that
both the operations of $a_{+}^{\dagger}(\vec{p})$ and $a_{-}(\vec{p})$
on the vacuum
increase the energy-momentum of the state by an amount $p^{\mu}$.
The interpretation of the positive energy particle state
$a_{+}^{\dagger}(\vec{p})$
is the usual one, namely,
one particle of energy-momentum $p^{\mu}$
is created on the empty vacuum.
Because annihilation of negative energy particles
corresponds with the creation of holes of positive energy,
the action of the annihilation operator $a_{-}(\vec{p})$
on the boson sea creates a bosonic hole of the momentum $p^{\mu}$.
The hole is just the anti-particle.
All other higher states are interpreted in the same manner.
Therefore,
the natural definition of the momentum operator $P^{\mu}$ in (\ref{4.14})
is consistent with our formulation.
We conclude that all the excited states, namely,
the particles and the holes have positive energies.

\section{Conclusion and future perspectives}

We have proposed a consistent formulation
of the {\it boson sea}
which is a bosonic version of the Dirac sea.
Our formulation shows that 
the boson vacuum forms a sea of bosons 
and the bosonic holes in the sea are interpreted as anti-particles.
To formulate the boson sea,
we have introduced the double harmonic oscillator,
the indefinite metric,
and a new definition of the inner product.
The non-local approach is employed
to define the positive definite inner product.
The double harmonic oscillator allows us to consider
the negative number states,
and is essential to realize the boson sea.
The negative energy solution of the field equation
belongs to the negative number sector,
then the bosonic hole in the boson sea always has positive energy.
We summarize an analogy between the Dirac sea and the boson sea
in Table~\ref{analogy}.

\begin{table}[htbp]
 \begin{center}
  \begin{tabular}{|c|c|c|} \hline
	& \multicolumn{2}{|c|}{\bf Fermions}\\\cline{2-3}
	& \parbox{12.5em}{\hfil Positive energy solutions\hfil}
	& \parbox{12.5em}{\hfil Negative energy solutions\hfil}\\
	& $E>0$ & $E<0$\\\hline
	& & \\[-0.5em]
	\parbox{12em}
	{\hfil Empty vacuum $||\tilde{0}_+\rangle$ of\hfil\\
	\hfil the positive number states\hfil}
	& \bf Realized in nature & Non-realizable in nature\\[-0.5em]
	& & \\ \hline
	& & \\[-0.5em]
	\parbox{12em}
	{\hfil Filled vacuum $||\tilde{0}_-\rangle$ of\hfil\\
	 \hfil the negative number states\hfil}
	& Non-realizable in nature
	& \parbox{10em}{\hfil{\bf Realized in nature}\hfil\\
			\hfil as the Dirac sea\hfil}
	\\[-0.5em]
	& & \\ \hline
	\multicolumn{3}{c}{} \\ \hline
	& \multicolumn{2}{|c|}{\bf Bosons}\\\cline{2-3}
	& Positive energy solutions
	& Negative energy solutions\\
	& $E>0$ & $E<0$\\\hline
	& & \\[-0.5em]
	\parbox{12em}
	{\hfil Empty vacuum $||{0}_+\rangle$ of\hfil\\
	\hfil the positive number states\hfil}
	& \bf Realized in nature & Non-realizable in nature\\[-0.5em]
	& & {\tiny (including solution of Klein-Gordon eq.)}\\\hline
	& & \\[-0.5em]
	\parbox{12em}
	{\hfil Filled vacuum $||{0}_-\rangle$ of \hfil\\
	 \hfil the negative number states \hfil}
	& Non-realizable in nature
	& \parbox{10em}{\hfil{\bf Realized in nature}\hfil\\
			\hfil as the boson sea\hfil}
	\\[-0.5em]
	& & \\\hline
  \end{tabular}
 \end{center}
 \caption{Analogy between fermion and boson sea}
 \label{analogy}
\end{table}

Supersymmetry has played an important role
to develop our method as the guiding principle of the boson sea.
Therefore,
our method is also natural when we consider supersymmetry.
Our method treats bosons and fermions on an equal footing
as the seas.

The concept of the boson sea is
widely applicable to quantum physics.
The string theories and the string field theories have been
successfully quantized only
by the light-cone quantization method~\cite{Kaku:1974zz}.
However,
there are no satisfactory theories of the covariant quantization
of the string field at the present time.
The first quantization of the string theory is performed
by the commutation relation
for the world sheet coordinates $X^{\mu}(\tau ,\sigma)$ as
\begin{eqnarray}
	&
	[X^{\mu}(\tau ,\sigma ),\Pi^{\nu}(\tau ,\sigma^{\prime})]
	=i\eta^{\mu \nu}\delta (\sigma -\sigma^{\prime}),
	&\label{string-com}\\
	&
	\quad \eta^{\mu \nu}={\rm diag}(-1,+1,\cdots ,+1).
	&\nonumber
\end{eqnarray}
This commutator has a negative sign for $\mu=\nu=0$
due to the metric element $\eta^{00}$ of the target space.
This property qualitatively coincides with
the commutation relation (\ref{B-com})
for the theory of the complex scalar field.
In the light-cone quantization,
the essential point of success is the disappearance of the
the negative energy states
caused by this negative sign of the commutator.
We expect that the covariant quantization of the string field
may well be obtained by the proper treatment of the negative energy states
in the string.
Because our method allows us to consider negative energy states,
our method may be useful for this purpose.

In this paper, we have succeeded in constructing the boson sea
formalism for free field theories.
The way of constructing the perturbation theory for interacting fields
without external fields is almost same as that for ordinary field
theories.
In the perturbation theory of the ordinary field theory, the
perturbative expansion is described by sum of the many harmonic
oscillator system, and the harmonic oscillators are not deformed.
It is just the same situation with the perturbation theory in our
formalism.
In this sense, there arises no difference and no ambiguity among the
ordinary field theory and our formalism.
On the other hand, the situation is changed when we switch on the
external fields or consider the non-perturbative properties.
One of the most important cases is the anomaly.

Also in analogy with the intuitive understanding and derivation of the
chiral anomaly of the massless fermion as pair creation from the Dirac
sea~\cite{Nielsen:1983rb},
we may expect to obtain a new insight about boson anomaly
such as the conformal anomaly.
Furthermore there is a possibility that the boson propagator may be
modified by the effect of the boson sea.
When we consider interactions and external fields,
it is expected that the interactions amang a particle,
the boson sea and the external fields
modify the behavior of the particle.

Before closing the present article, an announcement is in order.
In this paper the inner product of
the system of the double harmonic oscillator
is defined in the non-local method as is presented in section 4,
which proves to be positive definiteness of the inner product.
However this non-local method is not only one way to provide
a positive definite inner product.
In the successive paper [arXiv:hep-th/0607182]
in our series of this subject,
we in fact show another method which is a kind
of $\varepsilon$-regularization and renormalization method,
and it also provides positive definite one.
The detail of this method and calculation are presented there.

\section*{Acknowledgements}

We acknowledge R.~Jackiw for his encouraging communication who shares
with us the view point that the Dirac sea method provides deep physical
understanding and intuition to some novel phenomena.
This work is supported by Grants-in-Aid for Scientific Research on
Priority Areas, Number of Areas 763, ``Dynamics of Strings and Fields'',
from the Ministry of Education of Culture, Sports, Science and
Technology, Japan.


\end{document}